\documentclass[intlimits,twoside,a4paper]{article}

\usepackage[cp1251]{inputenc}

\usepackage[eqsecnum]{cmpj3}

\allowdisplaybreaks[3]

\issue{2018}{21}{2}{23002}
\doinumber{10.5488/CMP.21.23002}

\title[BEC and modulation in the two-state Bose-Hubbard model]%
{Bose-Einstein condensation and/or modulation of
``displacements'' in the two-state Bose-Hubbard model}
\author{I.V. Stasyuk, O.V. Velychko}
\address{Institute for Condensed Matter Physics of the National
Academy of Sciences of Ukraine,\\
1 Svientsitskii St., 79011 Lviv, Ukraine}

\date{Received May 8, 2018}

\begin{document}

\maketitle

\begin{abstract}
Instabilities resulting in Bose-Einstein condensation
and/or modulation of ``displacements'' in a system of quantum
particles described by a two-state Bose-Hubbard model (with an allowance for the interaction between particle displacements on different lattice sites) are
investigated. A possibility of modulation, which doubles the lattice constant, as well as the uniform displacement of particles from equilibrium positions are studied. Conditions for realization of the mentioned instabilities and phase transitions into the SF phase and into the ``ordered'' phase with frozen displacements are analyzed. The behaviour of order parameters is investigated and phase diagrams of the system are calculated both analytically (ground
state) and numerically (at non-zero temperatures). It is revealed that the SF phase can appear as an intermediate one between the normal and ``ordered'' phases, while a supersolid phase is thermodynamically unstable and does not appear. The relation of the obtained results to the lattices with the double-well local potentials is discussed.
\pacs 03.75.Hh, 03.75.Lm, 64.70.Tg, 71.35.Lk, 37.10.Jk, 67.85.-d
\end{abstract}

\section{Introduction}

Physical properties of the system of ultracold bosonic atoms in optical lattices are intensively studied during the last two decades. Beside a study of thermodynamics and the phase transition to the superfluid (SF) phase for such lattices of different structure and dimensionality, collective phenomena caused by interactions of various type are another matter of interest. Theoretical description is based on the Bose-Hubbard model (BHM) proposed in works \cite{Fisher:89:546,Jaksch:98:3108} and extended by consideration of direct intersite interactions between particles (resulting in the appearance of  modulated and separated phases) \cite{Buechler:03:130404} as well as by taking into account the excited single-site states \cite{Stasyuk:11:13004}. The extension of a single-site basis that also takes place in the case of bosons with nonzero spin (e.g., $S=1$, \cite{Demler:02:163001,Krutitsky:04:063610}) leads to more complex phase diagrams where the phase transition to the superfluid phase can change its order from the second to the first one. However, this transition is dominated by the on-site correlation $U$ of bosons of Hubbard type responsible for a characteristic multilobe shape of phase diagrams $(t,\mu)$ \cite{Jaksch:98:3108,Buechler:03:130404,Stasyuk:11:13004,Demler:02:163001,Krutitsky:04:063610,Ohashi:06:033617}, where $t$ is an energy of the boson hopping between the neighbouring sites, and $\mu$ is the boson chemical potential.

In the limit $U\rightarrow\infty$ (corresponding to the case of the so-called hard-core bosons (HCB), i.e., no more than one boson per site, $n_{i}\leqslant1$), this theory is applicable to the region where the two above-mentioned lobes connect at $T=0$ \cite{Stasyuk:09:539}. Being rather some kind of approximation, the HCB model is still widely exploited for a description of the Bose-Einstein condensate (BEC) in optical lattices (see \cite{Sengupta:07:063625}). Beside ultracold systems, the model is used in the theory of ionic conductors \cite{Mahan:76:780}, the systems of particles  intercalated or adsorbed on the crystal surfaces \cite{Astaldi:92:90,Velychko:09:249,Mysakovych:10:228}, as well as in describing the local electron pairing in the theory of high-temperature superconductors \cite{Micnas:90:113}.

Within the framework of the HCB model extended by an allowance for the first excited local vibrational state, we studied a phase transition to the superfluid phase for the case of a particle transfer over the excited states \cite{Stasyuk:11:13004}. It was established that this phase transition can be of the first or of the second order; the system can also separate into normal (NO) and superfluid (SF) phases. The study was continued by consideration of  non-ergodic effects and their contribution to the momentum distribution of bosonic particles \cite{Stasyuk:12:33002}.

The above-mentioned two-state HCB model was extended by taking into account the displacements of bosonic particles from their equilibrium positions in the optical lattice as well as their intersite interaction leading to the excitations of a phonon type \cite{Stasyuk:15:43004}. These excitations are related to transitions between the ground and excited states of bosonic atoms on lattice sites. Calculations of the displacement correlator (the Green's function $\langle\langle\hat{x}|\hat{x}\rangle\rangle$) determining a static susceptibility of ``dipole'' type revealed a tendency to the appearance of soft phonon modes and related instabilities with respect to the lowering of an initial symmetry and the transition to the state with ``frozen'' displacements.

The present work is aimed at a detailed study of this problem. We will consider the coexistence and the mutual influence of the BEC and the modulation (or the uniform ordering) of the particle displacements. Our description is based on the two-state HCB model where the particle transfer occurs between the local ground states (unlike the excited-state transfer considered in \cite{Stasyuk:11:13004,Stasyuk:15:43004}). This approach takes into account the peculiarities of boson transfer in the optical lattice with local double-well potentials \cite{Albiez:05:010402,Danshita:07:043606}. Such a lattice is a valuable test bench allowing for an easy tuning of characteristic parameters of the model (e.g., the energy of tunnel splitting in double wells). We will study the thermodynamics of the described system calculating the phase diagrams and establishing the conditions for the phase coexistence. Possible applications of the obtained results beyond the optical lattices will be discussed as well.

\section{Model}

The Hamiltonian of the HCB system on a lattice, in which the two lowest local levels separated by an energy gap $\delta$ $(\delta>0)$ are considered only, can be written as
\begin{equation}
	\hat{H}=\hat{H}_{0}+\hat{H}'+\hat{H}'',
\label{istaeq1.1}
\end{equation}
where
\[
	\hat{H}_{0}
	=
	\sum_{ij} t_{ij} b_{i}^{+} b_{j}
	+
	\sum_{ij} t'_{ij} c_{i}^{+} c_{j}
	-
	\mu \sum_{i} b_{i}^{+} b_{i}
	+
	(\delta-\mu) \sum_{i} c^{+}_{i} c_{i}
\]
is the Hamiltonian of noninteracting bosons consisting of the terms that describe the intersite boson transfer over the ground ($t_{ij}$) and excited ($t'_{ij}$) states,
\begin{equation}
	\hat{H}' = \frac12\sum_{ij}\Phi_{ij}\hat{x}_{i}\hat{x}_{j}
\label{istaeq1.2}
\end{equation}
corresponds to the interaction between particle displacements (relative to their equilibrium positions in the respective lattice sites),
\begin{equation}
	\hat{H}'' = -h\sum_{i}\hat{x}_{i}
\label{istaeq1.3}
\end{equation}
describes the influence of the external field conjugated to the displacements. Here,
\begin{equation}
	\hat{x}_{i} = d(c_{i}^{+}b_{i}+b_{i}^{+}c_{i}),
\label{istaeq1.4}
\end{equation}
where $d$ is a matrix element of the coordinate operator between the ground and excited states, $b_{i}$ ($b_{i}^{+}$) and $c_{i}$ ($c_{i}^{+}$) are the bosonic destruction (creation) operators in the respective states.

The following considerations will be performed in the limit $t_{ij}'\rightarrow0$, i.e., the particles will transfer over the ground states only. Such an approach corresponds to the case of bosonic particles moving in the lattice formed by double-well potentials (see appendix~\ref{app-1}).

Due to the chosen HCB limit, the basis of local states $|n_{i}^{b},n_{i}^{c}\rangle$ is reduced to the three ($|0\rangle=|00\rangle$, $|1\rangle=|10\rangle$, $|2\rangle=|01\rangle$) ones \cite{Stasyuk:11:13004}. Thus, in the Hubbard projection operator (X-operator) representation
\begin{gather}
	b_{i} =	X_{i}^{01}, 
	\qquad
	c_{i}=X_{i}^{02}, 
	\qquad 
	\hat{x}_{i}=d\left(X_{i}^{21}+X_{i}^{12}\right);
	\notag\\
	b_{i}^{+}b_{i}=X_{i}^{11},
	\qquad 
	c_{i}^{+}c_{i}=X_{i}^{22}.
\label{istaeq1.5}
\end{gather}
This results in
\begin{align}
	\hat{H}
	&=
	\sum_{ij}t_{ij}'X_{i}^{10}X_{j}^{01}
	-
	\mu\sum_{i}X_{i}^{11}
	+
	(\delta-\mu)\sum_{i}X_{i}^{22}
	\notag\\
	&\quad
	{}+
	\frac12\sum_{ij}\Phi_{ij}d^{2}
		\left(X_{i}^{21}+X_{i}^{12}\right)\left(X_{j}^{21}+X_{j}^{12}\right)
	\notag\\
	&\quad
	{}-
	hd\sum_{i}\left(X_{i}^{21}+X_{i}^{12}\right)
	.
\label{istaeq1.6}
\end{align}

Our aim  is to study the thermodynamics and equilibrium states of the considered model as well to investigate the phase transitions related to the BEC appearance and the particle displacements from equilibrium positions. The boson transfer and the interaction $\Phi_{ij}$ are considered in the mean field approximation (MFA):
\begin{equation}
	\sum_{ij}t_{ij}X_{i}^{10}X_{j}^{01}
	\rightarrow 
	t(0)\xi \sum_{i}\left(X_{i}^{10}+X_{i}^{01}\right)-N t(0)\xi^{2},
\label{istaeq1.7}
\end{equation}
where
$t(0)=\sum_{j}t_{ij}$, 
$\xi=\left\langle X_{i}^{01}\right\rangle=\left\langle X_{i}^{10}\right\rangle$ (see \cite{Stasyuk:11:13004}), 
and
\begin{equation}
	\frac12\sum_{ij}\Phi_{ij}\hat{X}_{i}\hat{X}_{j}
	\rightarrow 
	\rho\Phi(0)d \sum_{i}\left(X_{i}^{12}+X_{i}^{21}\right)
	-
	\frac{N}{2} \Phi(0)\rho^{2},
\label{istaeq1.8}
\end{equation}
while 
$\Phi(0)=\sum_{j}\Phi_{ij}$, 
$\rho=\langle \hat{x}_{i}\rangle=2d\langle X_{i}^{12}\rangle$ 
(here $\langle X_{i}^{12}\rangle=\langle X_{i}^{21}\rangle$).

In this approximation
\begin{equation}
	\hat{H}_{\text{MF}}
	=
	\sum_{i}\hat{H}_{i}
	-
	{N} t(0) \xi^{2}-\frac{N}{2}\Phi(0)\rho^{2}
	,
\label{istaeq1.9}
\end{equation}
where
\begin{align}
	\hat{H}_{i}
	&= 
	t(0) \xi\left( X_{i}^{10}+X_{i}^{01}\right)
	-
	\mu X_{i}^{11}+(\delta-\mu)X_{i}^{22}
	\notag\\
	&\quad{}+
	\rho\Phi(0)d\left(X_{i}^{12}+X_{i}^{21}\right)
		-
		hd\left(X_{i}^{12}+X_{i}^{21}\right)
	.
\label{istaeq1.10}
\end{align}
Now energies of local bosonic states are determined by the eigenvalues of $\lambda_{\alpha}$ of the matrix 
\begin{equation}
    ||\hat{H}_{i}||
    =
    \left(
    \begin{array}{ccc}
		0		&	t(0)\xi				&	0 \\
		t(0)\xi	&	-\mu				&	[\rho\Phi(0)-h]d\\
		0		&	[\rho\Phi(0)-h]d	& 	\delta-\mu \\
    \end{array}
    \right)
    .
\label{istaeq1.11}
\end{equation}
Correspondingly, the grand canonical potential of the model looks like
\begin{equation}
	\Omega/N
	=
	-t(0)\xi^{2}-\frac12\Phi(0)\rho^{2}
%	\notag\\
	-
	\Theta 
	\ln
	\left(
		\rm e^{-\beta \lambda_{1}}
		+
		\rm e^{-\beta \lambda_{2}}
		+
		\rm e^{-\beta \lambda_{3}}
	\right)
	.
\label{istaeq1.12}
\end{equation}

Order parameters $\xi$ and $\rho$ are established from the self-consistency equations
\begin{equation}
	\frac{\partial\Omega/N}{\partial\xi}=0,
	\qquad 
	\frac{\partial\Omega/N}{\partial\rho}=0.
\label{istaeq1.13}
\end{equation}
They correspond to the absolute minimum of the grand potential at certain values of the chemical potential~$\mu$ and the field $h$.

In principle, there are phases with $\rho=0$ and $\xi\neq0$, $\rho\neq0$ and $\xi=0$ as well as $\rho\neq0$ and $\xi\neq0$ (apart from the normal one with $\rho=0$ and $\xi=0$). At first, we will consider the transitions to the phases with $\rho\neq0$ and $\xi\neq0$ as independent ones.

\section[Phase with $\rho\neq0$]%
{\mathversion{bold}%
Phase with $\rho\neq0$}

The solution of the equation system \eqref{istaeq1.13} with  $\rho\neq0$ and $\xi=0$ describes a phase with uniform spontaneous displacements of particles. In the case of the lattice with local potentials as the double wells, such a solution corresponds to an asymmetric occupation of minima of the well (like a uniform dipole ordering). At $\xi=0$, the eigenvalues of the matrix \eqref{istaeq1.11} are
\begin{equation}
	\lambda_{1}=0,
	\quad 
	\lambda_{2,3}=\frac{\delta}{2}-\mu\pm\sqrt{\delta^{2}/4+B^{2}}
	,
\label{istaeq2.1}
\end{equation}
where $B=d[\rho\Phi(0)-h]$.

The grand canonical potential is
\begin{equation}
	\Omega/N
	=
	\frac{1}{2}\Phi(0)\rho^{2}
	-
	\Theta\ln
		\left[
			1+2\cosh\beta\sqrt{\delta^{2}/4+B^{2}}{\rm e}^{-\beta(\delta/2-\mu)}
		\right]
	,
\label{istaeq2.2}
\end{equation}
the $\rho$ parameter is determined by the second equation from the set  \eqref{istaeq1.13}
\begin{equation}
	\rho
	=
	-
	\frac{2\sinh \beta R\cdot {\rm e}^{-\beta(\delta/2-\mu)}}
		{1+2\cosh \beta R\cdot {\rm e}^{-\beta(\delta/2-\mu)}}
	\cdot
	\frac{B d}{R}
	\,,
\label{istaeq2.3}
\end{equation}
where the notation $R=\sqrt{\delta^{2}/4+B^{2}}$ is introduced.

At the non-zero field $h$, there is no a trivial solution $\rho=0$ of equation \eqref{istaeq2.3}. Such a solution exists only at $h=0$, when other solutions are determined by the equation
\begin{equation}
	1
	=
	-
	\frac{2\sinh \beta R_{0}\cdot {\rm e}^{-\beta(\delta/2-\mu)}}
		{1+2\cosh \beta R_{0}\cdot {\rm e}^{-\beta(\delta/2-\mu)}}
	\cdot
	\frac{d^{2}\Phi(0)}{R_{0}}
	\,,
\label{istaeq2.4}
\end{equation}
where $R_{0}=\sqrt{\delta^{2}/4+d^{2}\Phi^{2}(0)\rho^{2}}$.

As one can see, this equation can have solutions for $\rho$ at $\Phi(0)<0$ only. If a condition $\Phi(0)>0$ holds, the uniform ordering of displacements $\langle \hat{x}_{0}\rangle$ changes into a modulated one. At the interaction limited to the nearest neighbours and to the modulation doubling, a lattice constant (when $\langle \hat{x}_{i}\rangle$ is equal to $+\rho$ or $-\rho$ for the first or second sublattices, respectively) an equation analogous to \eqref{istaeq2.4} has a different sign of the right-hand side (see appendix~\ref{app-2}).

Thus, we can consider a single equation
\begin{equation}
	1
	=
	\frac{2\sinh \beta R_{0}\cdot {\rm e}^{-\beta(\delta/2-\mu)}}
		{1+2\cosh \beta R_{0}\cdot {\rm e}^{-\beta(\delta/2-\mu)}}
	\cdot
	\frac{d^{2}|\Phi(0)|}{R_{0}}
	\,,
\label{istaeq2.5}
\end{equation}
which is valid for the both cases.

Solutions of equation \eqref{istaeq2.5} are calculated numerically, choosing the ones corresponding to the absolute minimum of function \eqref{istaeq2.2}. In the limit of the zero temperature, this task greatly simplifies and is carried out analytically.

\subsection[Case $T=0$]%
{\mathversion{bold}%
Case $T=0$}

At the absolute zero temperature, only the ground state of the system contributes to  thermodynamic potential. Depending on the relations between the model parameters, only the single-site states with energies $\lambda_{1}$ and $\lambda_{3}$ can be the ground ones. In the first case
\begin{equation}
	\Omega/N
	=
	-\frac{1}{2}\Phi(0)\rho^{2}+\lambda_{1}
	=
	-\frac12\Phi(0)\rho^{2}
\label{istaeq2.6}
\end{equation}
and the equality $\rho=0$ follows from conditions \eqref{istaeq1.13}. In this phase, the particles (and, of course, their displacements) are absent.

When the ground state corresponds to the level $\lambda_{3}$,
\begin{equation}
	\Omega/N
	=
	-\frac{1}{2}\Phi(0)\rho^{2}+\delta/2-\mu-\sqrt{\delta^{2}/4+B^{2}}
	.
\label{istaeq2.7}
\end{equation}
Based on conditions \eqref{istaeq1.13}, we obtain the following equation for the parameter $\rho$
\begin{equation}
	\rho=-\frac{B d}{\sqrt{\delta^{2}/4+B^{2}}}
	.
\label{istaeq2.8}
\end{equation}
After replacement $\Phi(0)=-|\Phi(0)|$ [for $\Phi(0)<0$], this equation can be rewritten as
\begin{equation}
	\rho
	=
	\frac{\bar{B} d}{\sqrt{\delta^{2}/4+\bar{B}^{2}}}
	\,,
\label{istaeq2.9}
\end{equation}
where $\bar{B}=d[\rho|\Phi(0)|+h]$. As stated above, the equation also holds in the case $\Phi(0)>0$.

At certain conditions, the ground state can be changed. Based on the equality $\lambda_{1}=\lambda_{3}$, one can conclude that it takes place at
\begin{equation}
	\rho
	=
	\frac{\sqrt{\mu(\mu-\delta)}}{|\Phi(0)|d}-\frac{h}{|\Phi(0)|}
	.
\label{istaeq2.10}
\end{equation}
It follows from formula \eqref{istaeq2.9} that $\rho\rightarrow\pm d$ at $h\rightarrow\pm\infty$, respectively. On the other hand, the solution of equation \eqref{istaeq2.9} with respect to $h$ is as follows:
\begin{equation}
	h
	=
	\frac{\delta}{2d}\cdot\frac{\rho}{\sqrt{d^{2}-\rho^{2}}}-\rho|\Phi(0)|
	.
\label{istaeq2.11}
\end{equation}
Obviously, at $\rho \ll d$
\begin{equation}
	h
	\approx
	\left[\frac{\delta}{2d^{2}}-|\Phi(0)|\right]\rho
	.
\label{istaeq2.12}
\end{equation}
It follows from the above expression that the curve of $\rho$ as a function of $h$ is $S$-shaped at
\begin{equation}
	\delta/2<d^{2}|\Phi(0)|\equiv W
	,
\label{istaeq2.13}
\end{equation}
defining a condition for the appearance of a spontaneous displacement at $h=0$
\begin{equation}
	\rho_{0}^{\text s}=d\sqrt{1-\frac{\delta^{2}}{4W^{2}}}
	\,,
\label{istaeq2.14}
\end{equation}
which can be obtained from \eqref{istaeq2.9}. It describes a state with a  uniform or modulated displacement [determined by the sign of $\Phi(0)$] of bosonic particles residing in the lattice sites. It is important that $\rho_{0}^{\text s}$ is independent of the chemical potential of bosons.

Thus, there are two possible phases of the system at $T=0$: (1) the phase with $\rho=0$ (the ground level is $\lambda_{1}$) and (2) the phase with $\rho=\rho_{0}^{\text s}$ (the ground level is $\lambda_{3}$). The transition between them can be investigated by calculation of the grand canonical potential $\Omega$ at variation of $\mu$ with the constraint $\lambda_{1}=\lambda_{3}$.

\begin{figure}[!b]
\centerline{\includegraphics[width=0.6\textwidth]{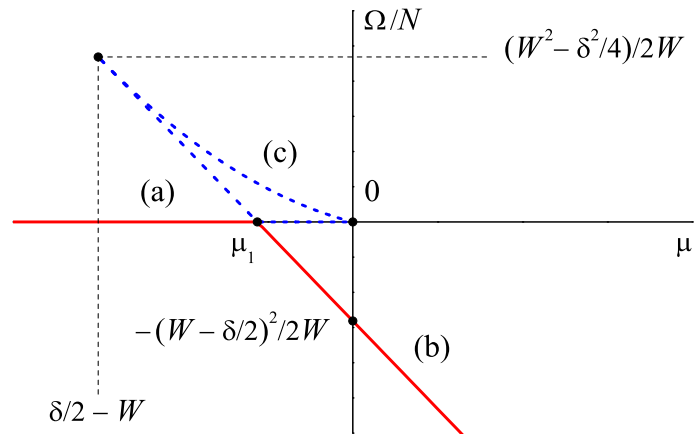}}
\caption{(Colour online) Dependence of the grand canonical potential $\Omega/N$ on the chemical potential of bosons $\mu$ at $T=0$ and $t(0)=0$ in the vicinity of a transition to the phase $\rho\neq0$. A phase transition of the first order occurs at $\mu=\mu_1$. Hereinafter, solid and dashed lines designate stable and metastable states, respectively.}
\label{fig01}
\end{figure}

Substituting $\rho$ [formula \eqref{istaeq2.10}] into the expression 
\begin{equation}
	\Omega/N
	=
	+\frac12|\Phi(0)|\rho^{2}+\lambda_{1}=-\frac12|\Phi(0)|\rho^{2}+\lambda_{3}
	\,,
\label{istaeq2.15}
\end{equation}
we obtain
\begin{equation}
	\Omega/N
	=
	\frac{1}{2|\Phi(0)|}
	\left[\frac{\sqrt{\mu(\mu-\delta)}}{d}+h\right]^{2}
	.
\label{istaeq2.16}
\end{equation}
In particular, at $h=0$
\begin{equation}
	\Omega/N=\frac{{\mu(\mu-\delta)}}{2W}
	.
\label{istaeq2.17}
\end{equation}
Plots of functions
\begin{subequations}
\label{istaeq2.18}
\begin{align}
	\Omega/N\big|_{\lambda_{1},\,h=0}
	&=
	0,
	\label{istaeq2.18a}
	\\
	\Omega/N\big|_{\lambda_{3},\,h=0}
	&=
	-\frac{(W-\delta/2)^{2}}{2W}-\mu,
	\label{istaeq2.18b}
	\\
	\Omega/N\big|_{\lambda_{1}=\lambda_{3},\,h=0}
	&=
	\frac{\mu(\mu-\delta)}{2W}
	\label{istaeq2.18c}
\end{align}
\end{subequations}
are presented in figure~\ref{fig01}. The fishtail shape of the dependence $\Omega(\mu)$ indicates the first order phase transition between phases  $\rho=0$ [branch \eqref{istaeq2.18a}] and $\rho\neq0$ [branch \eqref{istaeq2.18b}]. The respective points of absolute instability are at $\mu=\delta/2-W$ and $\mu=0$. Branch \eqref{istaeq2.18c} describes an unstable state. The phase transition occurs at
\begin{equation}
	\mu
	=
	\mu_{1}\equiv-\frac{1}{2W}\left(W-\frac{\delta}{2}\right)^{2}.
\label{istaeq2.19}
\end{equation}

\begin{figure}[!b]
\centerline{\includegraphics[width=0.6\textwidth]{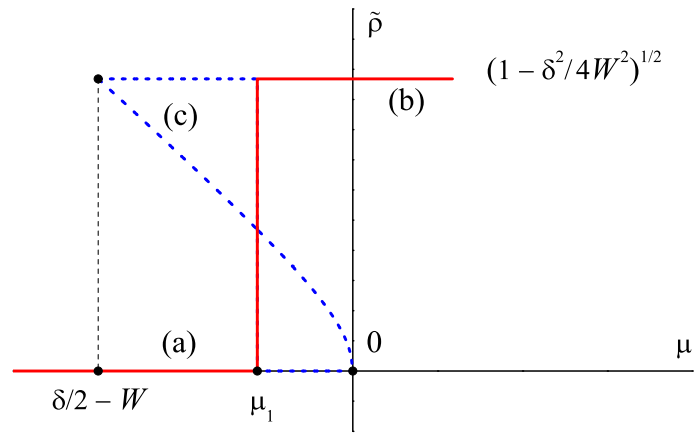}}
\caption{(Colour online) Dependence of the order parameter $\rho$ on the chemical potential of bosons $\mu$ at $T=0$ and $t(0)=0$ in the vicinity of the transition to the phase $\rho\neq0$.}
\label{fig02}
\end{figure}

The dependence of the order parameter 
$\rho \equiv \tilde{\rho} d$ 
on the chemical potential of bosons is described at $h=0$ by formulae
\begin{subequations}
\label{istaeq2.20}
\begin{align}
	\tilde{\rho}
	&=
	0,
	\label{istaeq2.20a}
	\\
	\tilde{\rho}
	&=
	\sqrt{1-\frac{\delta^{2}}{4W^{2}}}\,,
	\label{istaeq2.20b}
	\\
	\tilde{\rho}
	&=
	\frac{\sqrt{\mu(\mu-\delta)}}{W}
	\label{istaeq2.20c}
\end{align}
\end{subequations}
and is presented in figure~\ref{fig02}. There is a jump between the values $\tilde{\rho}=0$ and $\tilde{\rho}=\rho_{0}^{\text s}/d$ at $\mu=\mu_{1}$.

This phase transition is accomplished by a jump of the boson concentration  $\bar{n}_{\text B}=-\partial(\Omega/N)/\partial\mu$ when $\bar{n}_{\text B}$ changes from $\bar{n}_{\text B}=0$ to $\bar{n}_{\text B}=1$ at the rise of $\mu$ [as it follows from formulae \eqref{istaeq2.6} and \eqref{istaeq2.7}].

\subsection[Case $T\neq0$]%
{\mathversion{bold}%
Case $T\neq0$}

The above considerations are limited to the zero temperature. At $T\neq0$, one should start from the equations \eqref{istaeq2.3} or \eqref{istaeq2.5} as well as from the general expression \eqref{istaeq2.2} for the grand canonical potential $\Omega$ and perform numerical calculations. As usual, a spontaneous (i.e., at $h=0$)  value of the order parameter $\rho$ decreases at the rise of temperature. The phase transition to the NO phase is of the second order. It occurs at the temperature defined by the equation
\begin{equation}
	1
	=
	\frac{2\sinh \frac{\beta\delta}{2} {\rm e}^{-\beta(\delta/2-\mu)}}%
		{1+2\cosh \frac{\beta\delta}{2} {\rm e}^{-\beta(\delta/2-\mu)}}
	\cdot
	\frac{2W}{\delta}\,,
\label{istaeq2.21}
\end{equation}
which follows from \eqref{istaeq2.5} in the limit $\rho\rightarrow0$.

\begin{figure}[!t]
\includegraphics[width=0.49\textwidth]{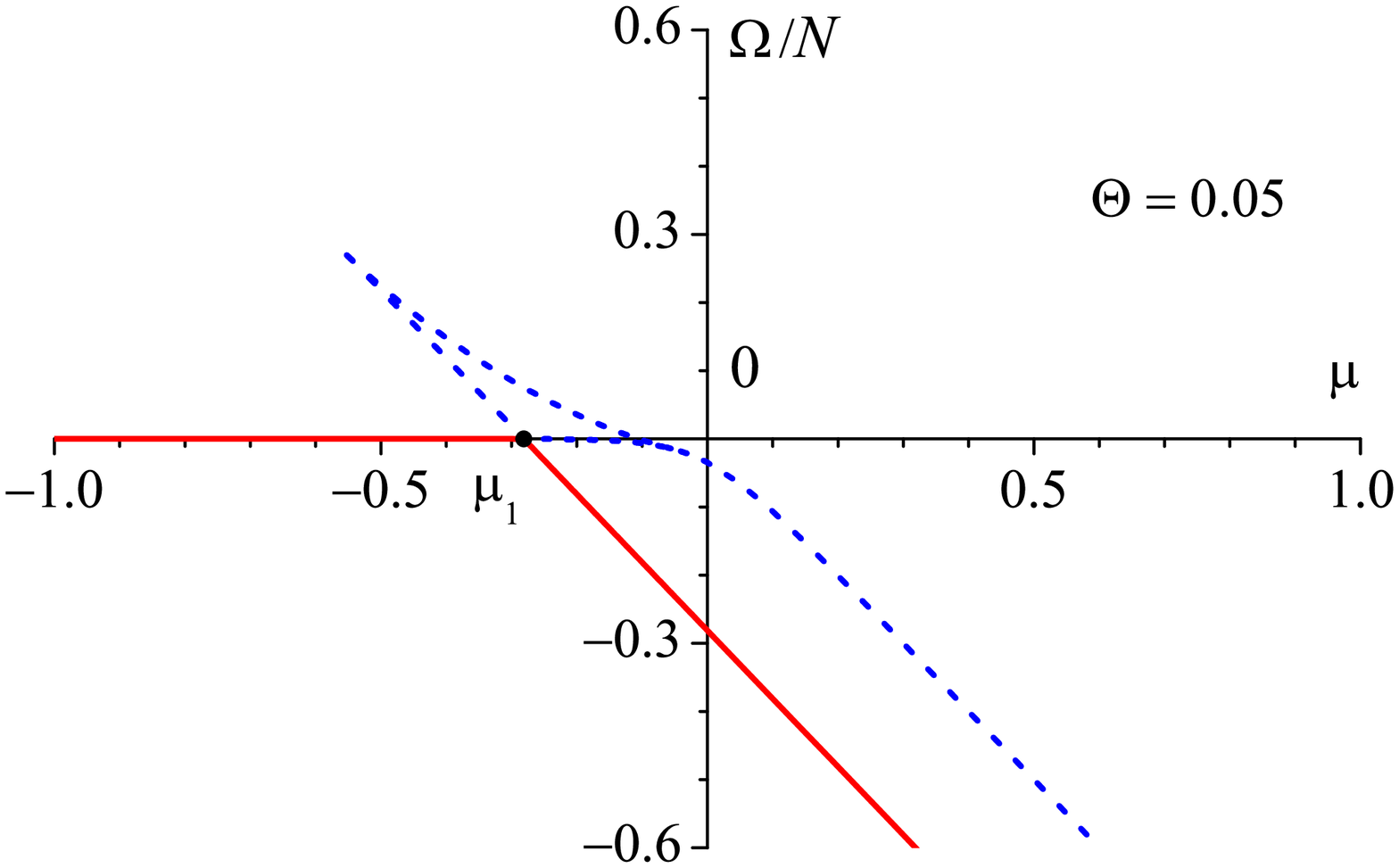}%
\hfill%
\includegraphics[width=0.49\textwidth]{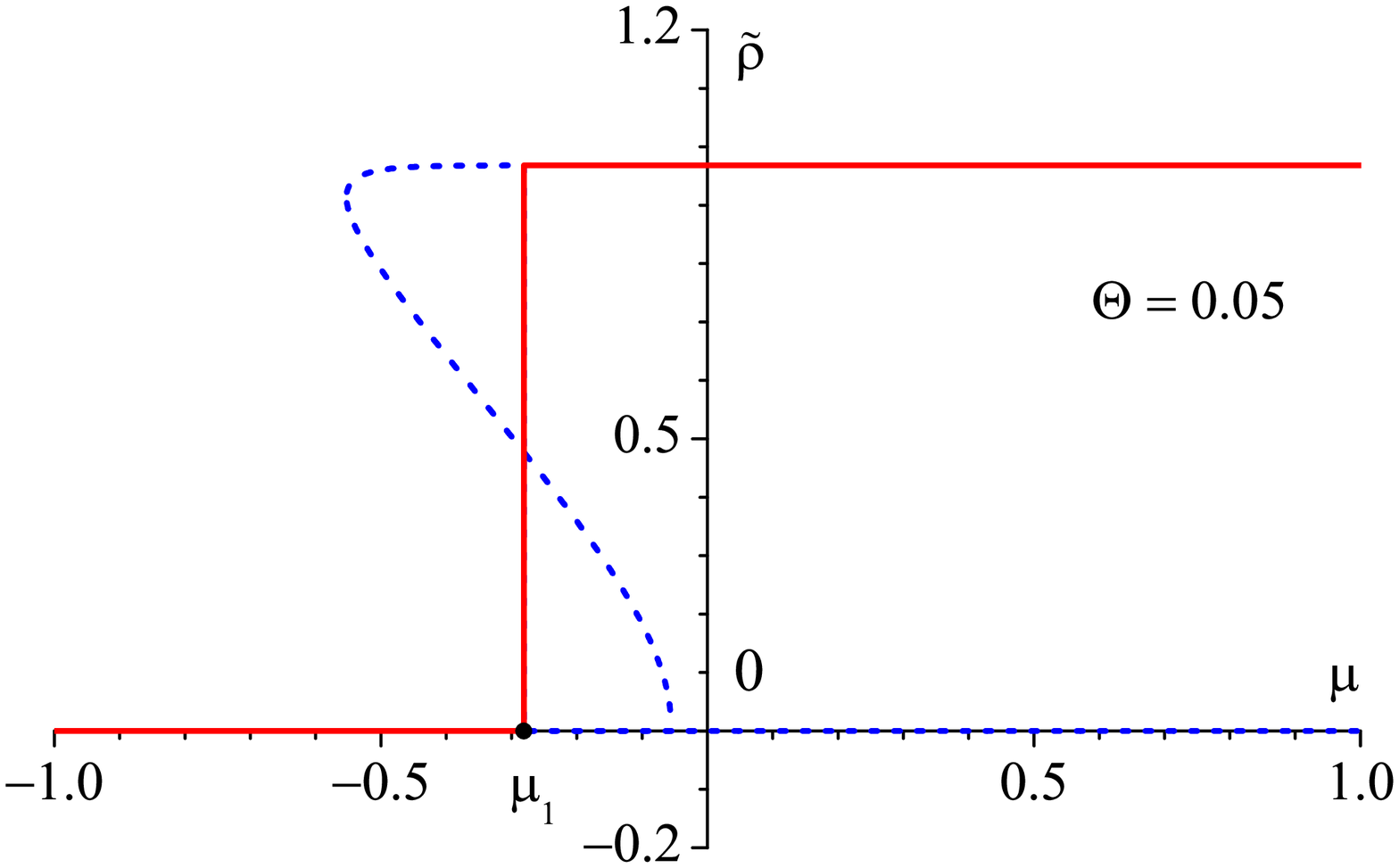}%
\\ [1.5ex]
\includegraphics[width=0.49\textwidth]{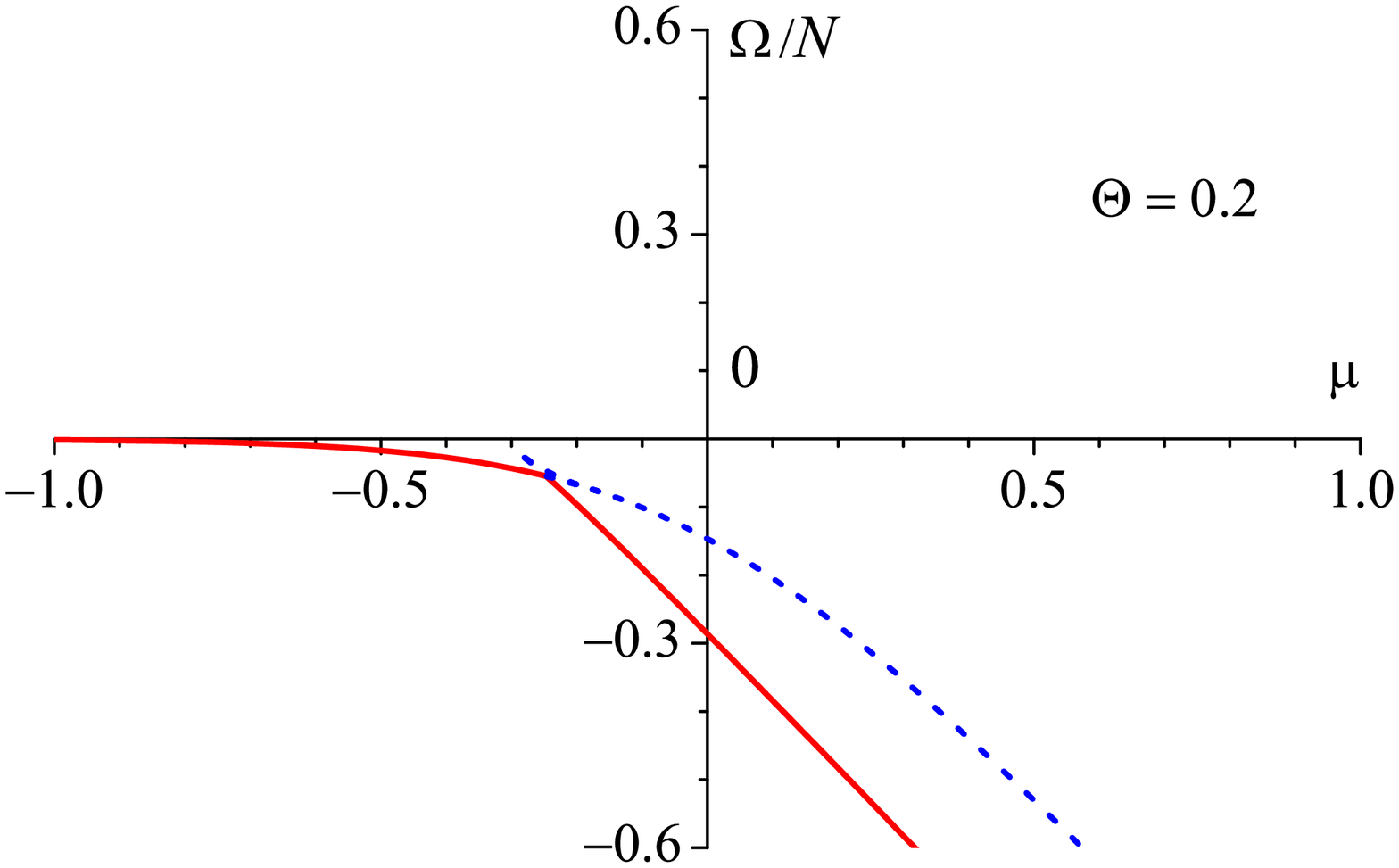}%
\hfill%
\includegraphics[width=0.49\textwidth]{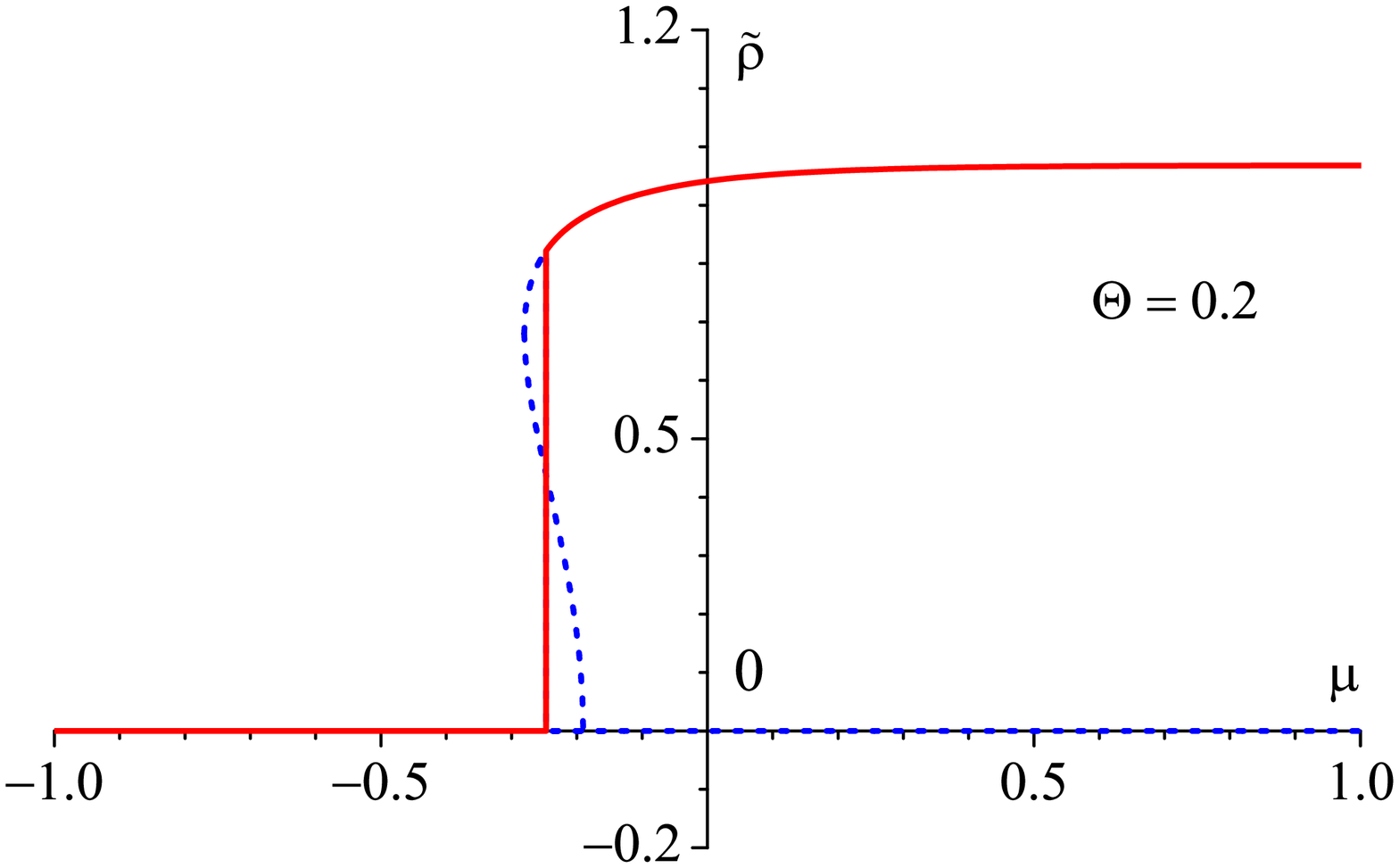}%
\\ [1.5ex]
\includegraphics[width=0.49\textwidth]{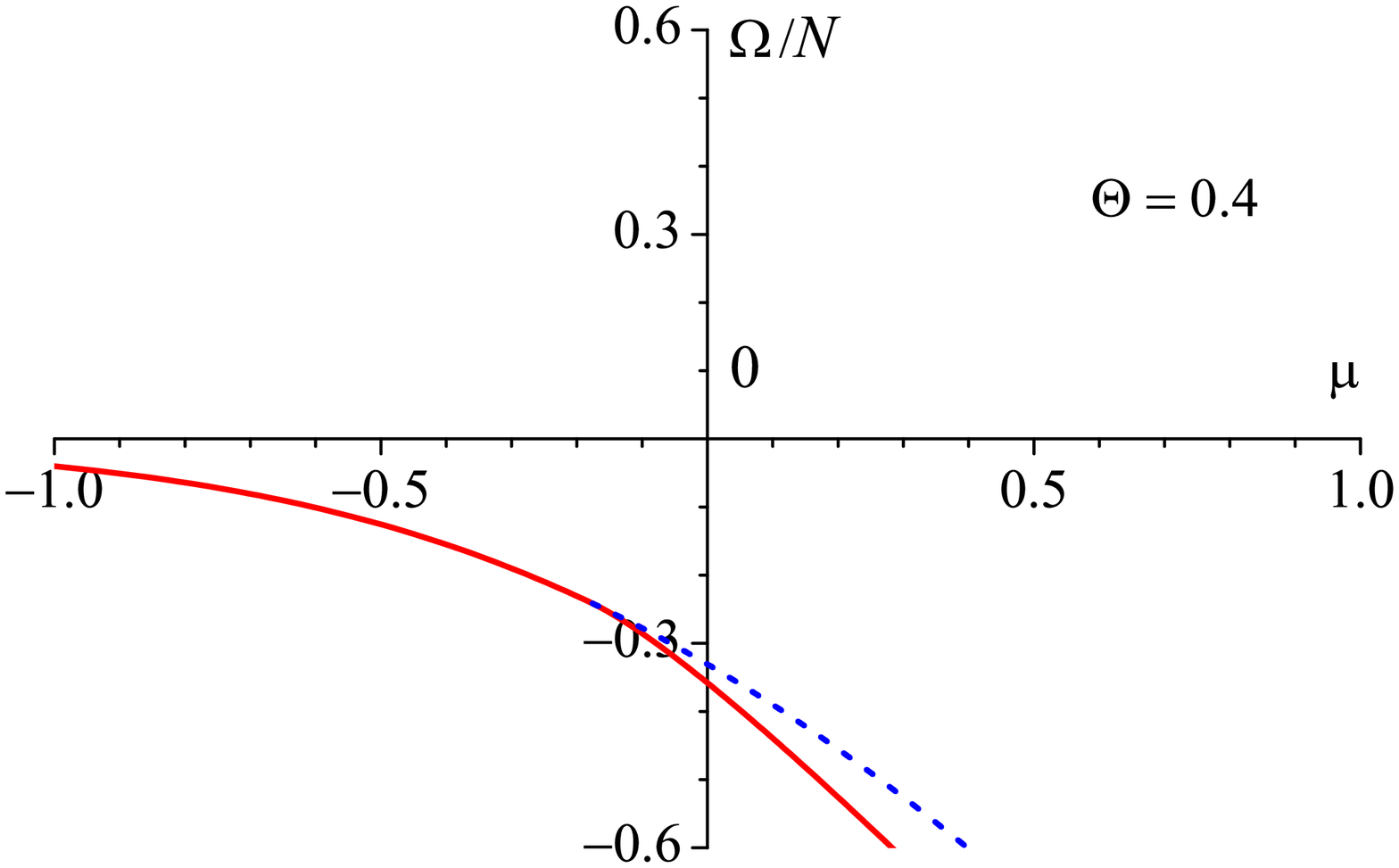}%
\hfill%
\includegraphics[width=0.49\textwidth]{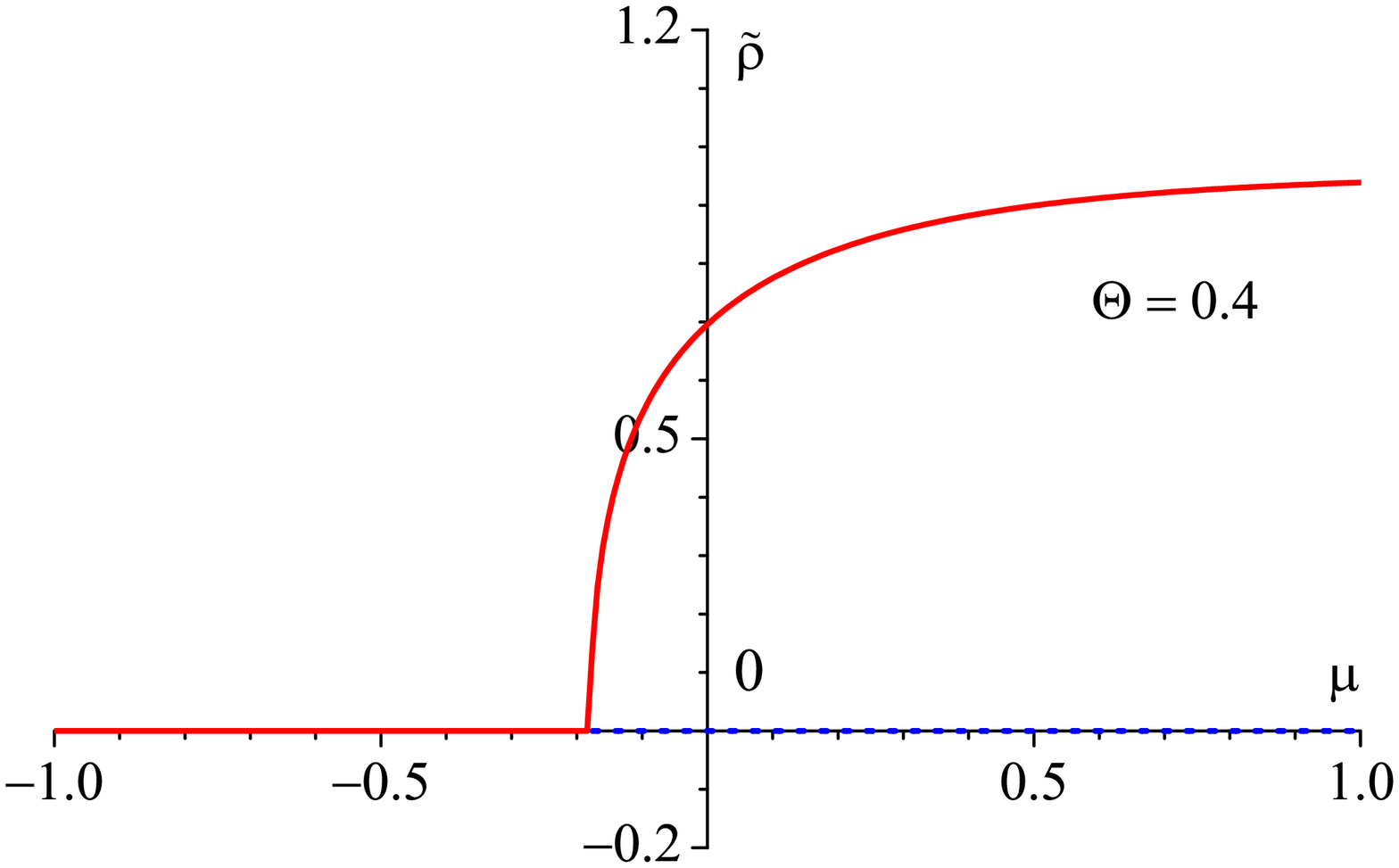}%
\caption{(Colour online) Dependences of the grand canonical potential $\Omega/N$ and the order parameter $\rho$ on the chemical potential of bosons $\mu$ at $T\neq0$ ($|t(0)|=0$, $\delta=0.5$ and $W=1$). Quantities having a dimension of energy are given in units of $W$.}
\label{fig03}
\end{figure}

\begin{figure}[!t]
\centerline{\includegraphics[width=0.6\textwidth]{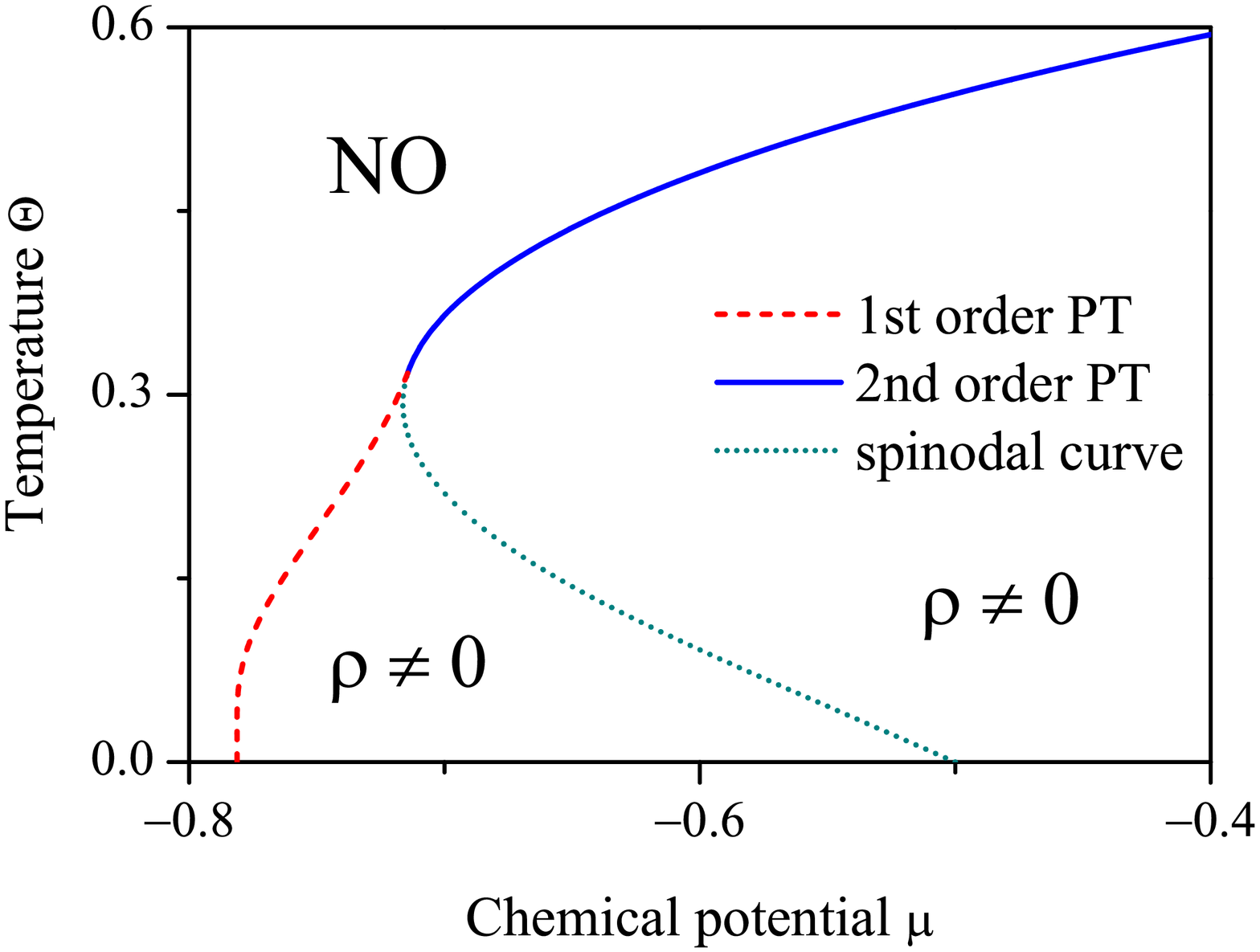}}
\caption{(Colour online) The phase diagram $(T,\mu)$ in the vicinity of the transition to the phase $\rho\neq0$ at $|t(0)|=0$, $\delta=0.5$ and $W=1$. Solid and dashed lines designate phase transition of the first and second order, respectively, while a dotted line corresponds to spinodal.}
\label{fig04}
\end{figure}

The dependences of $\Omega$ and $\rho$ on $\mu$ at various temperatures are depicted in figure~\ref{fig03}. The corresponding phase diagram $(T,\mu)$ shown in figure~\ref{fig04} defines the region of the $\rho\neq0$ phase at a certain relationship between the parameter $W$ (describing the interaction of bosonic displacements) and the energy $\delta$ of the excited bosonic state on a lattice site. Figure~\ref{fig04} also demonstrates the spinodal curve where the NO phase reaches an absolute instability [the spinodal is defined by the equality $(\partial\rho/\partial\mu)|_{\rho=0}=\infty$].

\section[Phase $\xi\neq0$]%
{\mathversion{bold}%
Phase $\xi\neq0$}

Here, we consider the phase transition between the NO and SF (with $\xi\neq0$) phases for the model~\eqref{istaeq1.1} (where bosons transfer through the ground state) assuming its independence of the phase transition with the order parameter $\rho$. Obviously, it happens in the absence of the interaction of displacements $\Phi_{ij}$. Later on, we will formulate the criterion for such a transition at non-zero values of $\Phi_{ij}$.

One can separate the part of the Hamiltonian which is related to the BEC. According to equations~\eqref{istaeq1.10} and \eqref{istaeq1.11}, in the MFA and under conditions $\rho=0$ and $h=0$, the single-site spectrum of bosons in the SF phase looks as follows:
\begin{equation}
	z_{1,2}
	=
	-\frac{\mu}{2}\pm\sqrt{\mu^{2}/4+t^{2}(0)\xi^{2}}
	,
	\qquad
	z_{3}
	= 
	\delta-\mu
	.
\label{istaeq3.1}
\end{equation}
In this case, the grand canonical potential $\Omega$ is expressed as
\begin{equation}
	\Omega/N
	=
	-t(0)\xi^{2}
	-\Theta\ln
		\left[
			{\rm e}^{-\beta(\delta-\mu)}
			+
			{\rm e}^{\frac{\beta\mu}{2}}
				\cdot
				\cosh \beta \sqrt{\frac{\mu^{2}}{4}+t^{2}(0)\xi^{2}}
		\,
		\right]
	.
\label{istaeq3.2}
\end{equation}
Thus, taking into account the equilibrium condition \eqref{istaeq1.13}, one can derive the following equation for the BEC order parameter
\begin{equation}
	\xi
	=
	-
	\frac{\sinh\beta\sqrt{\mu^{2}/4+t^{2}(0)\xi^{2}}}
		{{\rm e}^{-\beta\delta}
			{\rm e}^{\beta\mu/2}+2\cosh\beta\sqrt{\mu^{2}/4+t^{2}(0)\xi^{2}}}
	\cdot
	\frac{t(0)\xi}{\sqrt{\mu^{2}/4+t^{2}(0)\xi^{2}}}
	.
\label{istaeq3.3}
\end{equation}
The above equation is known from the two-state model. It was obtained in a similar form in \cite{Stasyuk:11:13004} for the case of the boson transfer through excited states, which corresponds to a different sign of the parameter~$\delta$.

Equation \eqref{istaeq3.3} has non-zero solutions at $t(0)<0$. At $\delta>0$, the phase transition to the SF phase is of the second order in the limit $T\rightarrow0$, as demonstrated in \cite{Stasyuk:11:13004}. This can be proved by analyzing  the ground state of a bosonic system: level $z_{2}$ is the lowest one for any value of $\mu$. For potential $\Omega$, we have
\begin{equation}
	\Omega/N
	=
	|t(0)|\xi^{2}
	-
	\mu/2
	-
	\sqrt{{\mu^{2}}/{4}+t^{2}(0)\xi^{2}}
	,
\label{istaeq3.4}
\end{equation}
while an equation for a non-zero $\xi$ is reduced to
\begin{equation}
	1=\frac{|t(0)|}{2\sqrt{{\mu^{2}}/{4}+t^{2}(0)\xi^{2}}}
	.
\label{istaeq3.5}
\end{equation}
Thus,
\begin{equation}
	\xi=\frac12\sqrt{1-\frac{\mu^{2}}{t^{2}(0)}}
	.
\label{istaeq3.6}
\end{equation}
The above expression demonstrates that the SF phase ($\xi\neq0$) exists in the region $-|t(0)|<\mu<|t(0)|$.

\begin{figure}[!t]
\centerline{\includegraphics[width=0.6\textwidth]{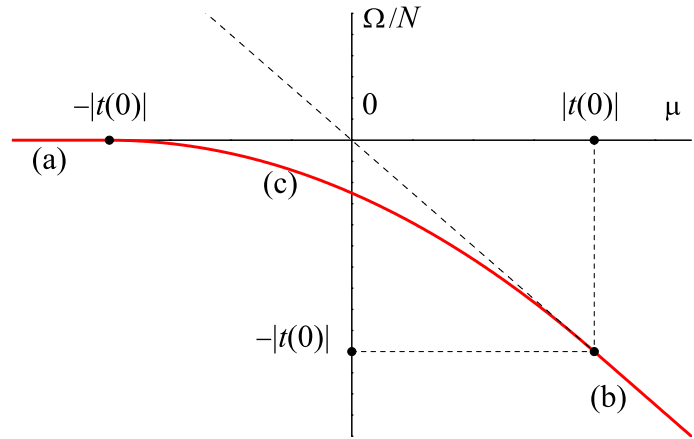}}
\caption{(Colour online) Dependence of the grand canonical potential $\Omega/N$ on the chemical potential of bosons $\mu$ in the vicinity of the transition to the SF phase ($\xi\neq0$) at $T=0$ ($\Phi_{ij}=0$).}
\label{fig05}
\end{figure}

\begin{figure}[!t]
\centerline{\includegraphics[width=0.6\textwidth]{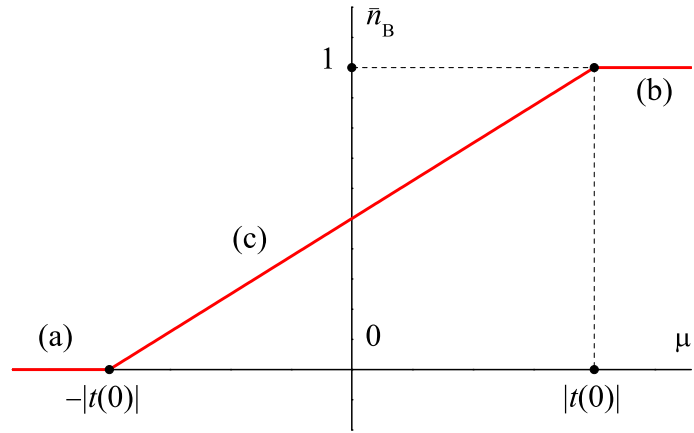}}
\caption{(Colour online) Dependence of the bosonic occupation number $\bar{n}_{\text{B}}$ on the chemical potential of bosons $\mu$ in the vicinity of the transition to the SF phase ($\xi\neq0$) at $T=0$ ($\Phi_{ij}=0$).}
\label{fig06}
\end{figure}

Substituting \eqref{istaeq3.6} into the expression for the grand canonical potential, one can obtain
\begin{equation}
	\Omega/N
	=
	\begin{cases}
		0,		&	\mu<-|t(0)|;	\\
		-\mu,	&	\mu>|t(0)|;		\\
		-\frac{1}{4|t_{0}|}[\mu+|t(0)|]^{2}, 
				&	-|t(0)|<\mu<|t(0)|.
	\end{cases}
\label{istaeq3.7}
\end{equation}
The derivative $\frac{\partial}{\partial\mu}\left(\Omega/N\right)=-\bar{n}_{\text B}$ (thus, defining the concentration of bosons) is continuous at points $\mu=\pm|t(0)|$. This is an evidence that the transition to the SF phase ($\xi\neq 0$) is of the second order (at $T=0$).
Dependences $\Omega/N$ and $\bar{n}_{\text B}$ on $\mu$ are presented in figures~\ref{fig05} and \ref{fig06}, respectively.

It should be mentioned that at $T\neq0$, due to a partial occupation of the excited state, the phase transition to the SF phase can change its order from the second to the first one. It takes place at low energies of excitation $\delta$ and at intermediate temperatures (this issue is discussed more in detail in \cite{Stasyuk:11:13004}).

\section{Complete phase diagrams}

All aforementioned conditions for the appearance of phases with $\xi\neq0$ or $\rho\neq0$ at $T=0$ are obtained assuming the phases as mutually exclusive. Such a conjecture is true at $\Phi_{ij}=0$ for the former phase and at $t_{ij}=0$ for the latter one. At non-zero parameters $\Phi_{ij}$ and $t_{ij}$, the domains of the above-mentioned phases intersect. Only the analysis of the grand potential $\Omega$ can detect the thermodynamically stable phases in this case. 

We have performed numerical calculations based on expression \eqref{istaeq1.12} with the use of eigenvalues of matrix \eqref{istaeq1.11} and self-consistency equation \eqref{istaeq1.13}. It is shown that only phases with $\xi\neq0$ or $\rho\neq0$ can be thermodynamically stable while the phase with the both non-zero order parameters ($\xi\neq0$ and $\rho\neq0$) is unstable at any conditions in the considered model. Thus, complete phase diagrams can be built by comparison of the grand canonical potentials for respective phases.

\begin{figure}[!t]
\includegraphics[width=0.48\textwidth]{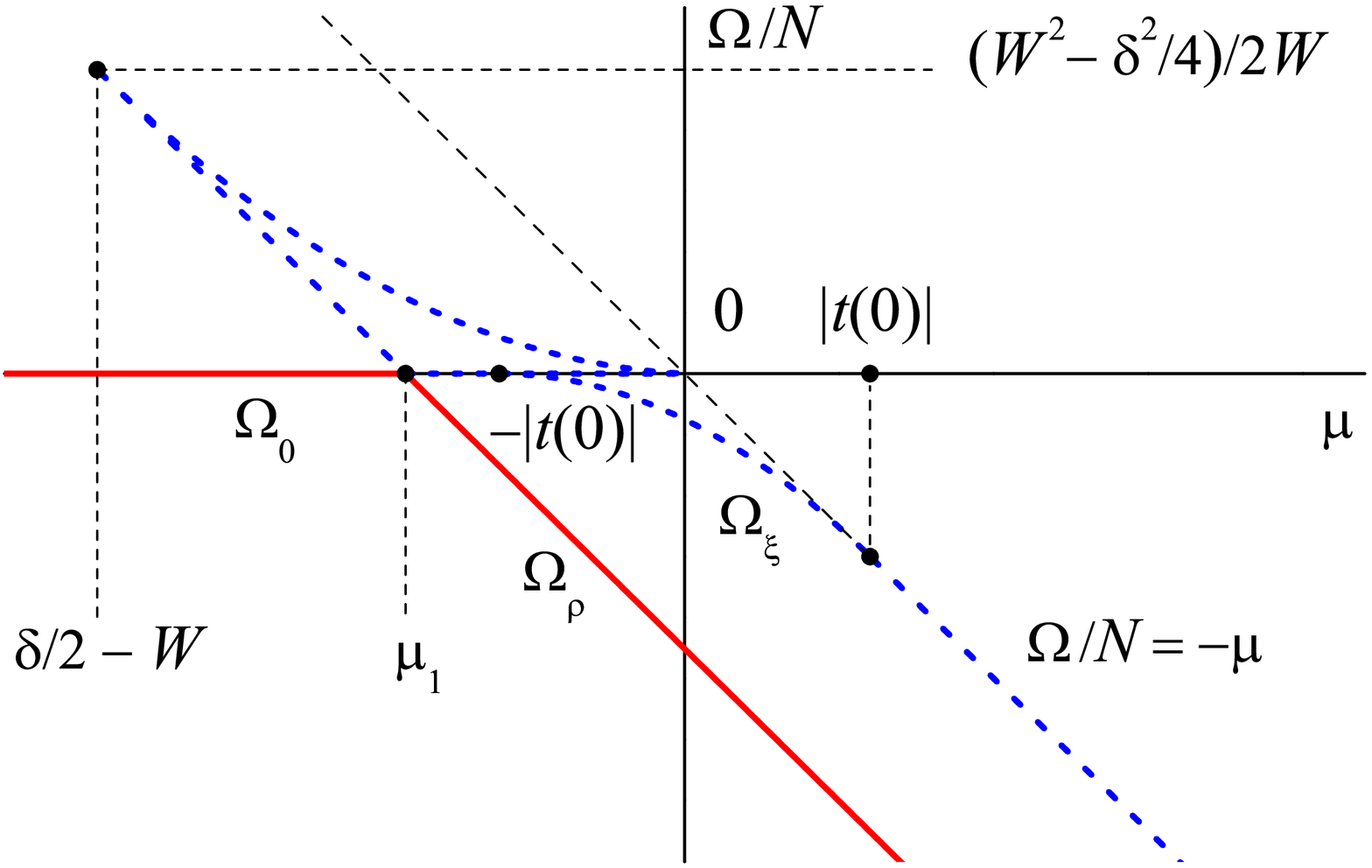}%
\hfill%
\includegraphics[width=0.48\textwidth]{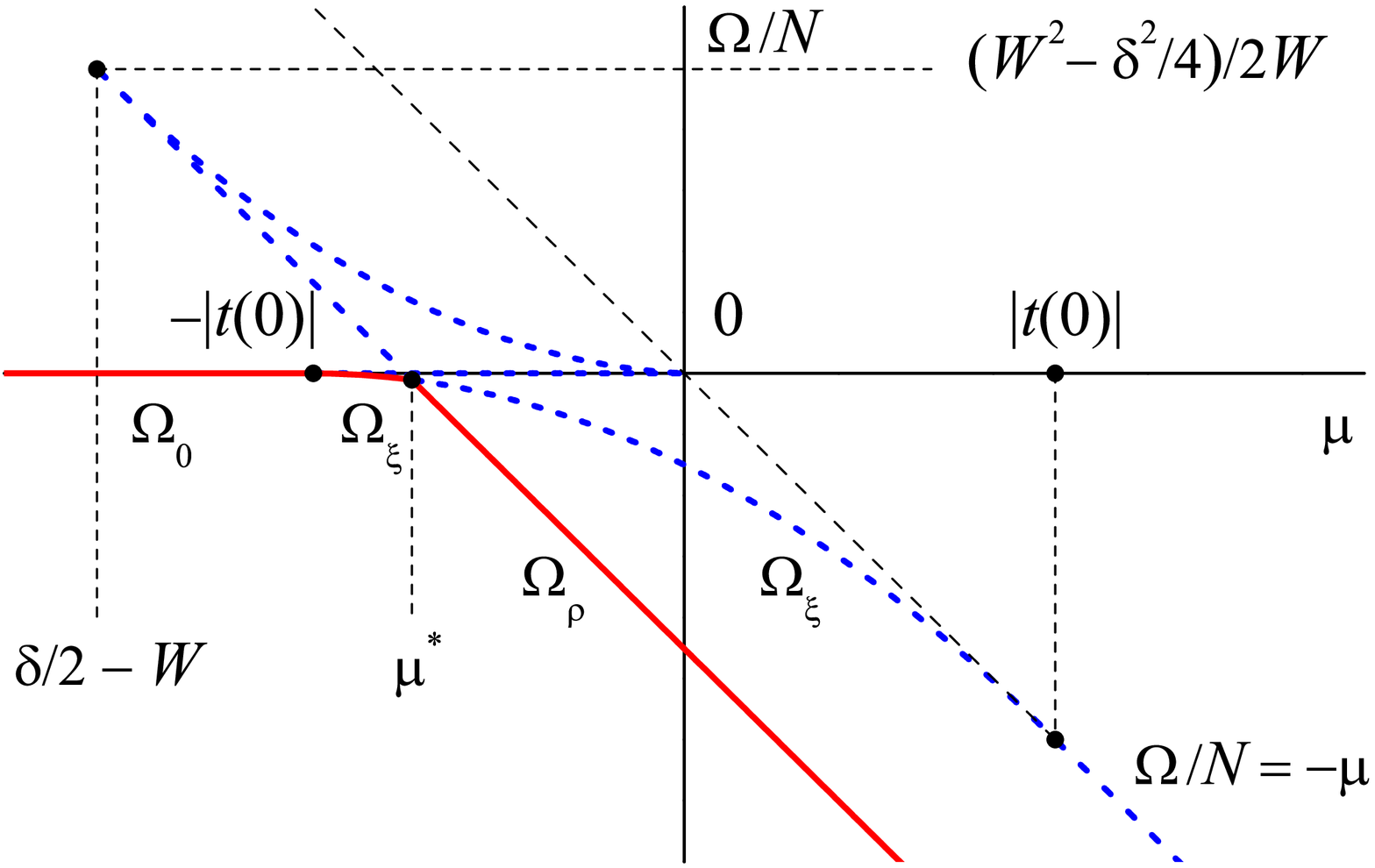}%
\caption{(Colour online) Grand canonical potentials $\Omega_{\rho}$ [formula~\eqref{istaeq2.18}] and $\Omega_{\xi}$ [formula~\eqref{istaeq3.7}] as functions of the chemical potential of bosons $\mu$ at $T=0$. On the left: a direct transition between the NO phase and the phase $\rho\neq0$ providing inequality~\eqref{istaeq4.1} is satisfied; on the right: the appearance of the intermediate SF phase ($\xi\neq0$) when this condition is broken.}
\label{fig07}
\end{figure}

At $T=0$, the consideration can be performed in an analytical form using formulae \eqref{istaeq2.18} and \eqref{istaeq3.7}. Plots of appropriate functions are combined in figure~\ref{fig07}, where two cases are separated. The first one [figure~\ref{fig07}~(a)] realizes at
\begin{equation}
	\frac{W}{2}\left(1-\frac{\delta}{2W}\right)^{2}>|t(0)|
	.
\label{istaeq4.1}
\end{equation}
The phase transition to the $\rho$ phase at an increase of $\mu$ takes place prior to the possible appearance of the BEC. The branch $\Omega_{\xi}$ (corresponding to the state with $\xi\neq0$) has greater values of $\Omega$ than the branch $\Omega_{\rho}$ (corresponding to the state with $\rho\neq0$); hence, the SF phase is unstable. Thus, there are sequentially two phases here: (1) the NO phase ($\rho=0$ and $\xi=0$) at $\mu<\mu_{1}$ and (2) the phase with $\rho\neq0$ at $\mu>\mu_{1}$
[$\mu_{1}=-W(1-\delta/2W)^{2}/2$].

\begin{figure}[!t]
\includegraphics[width=0.48\textwidth]{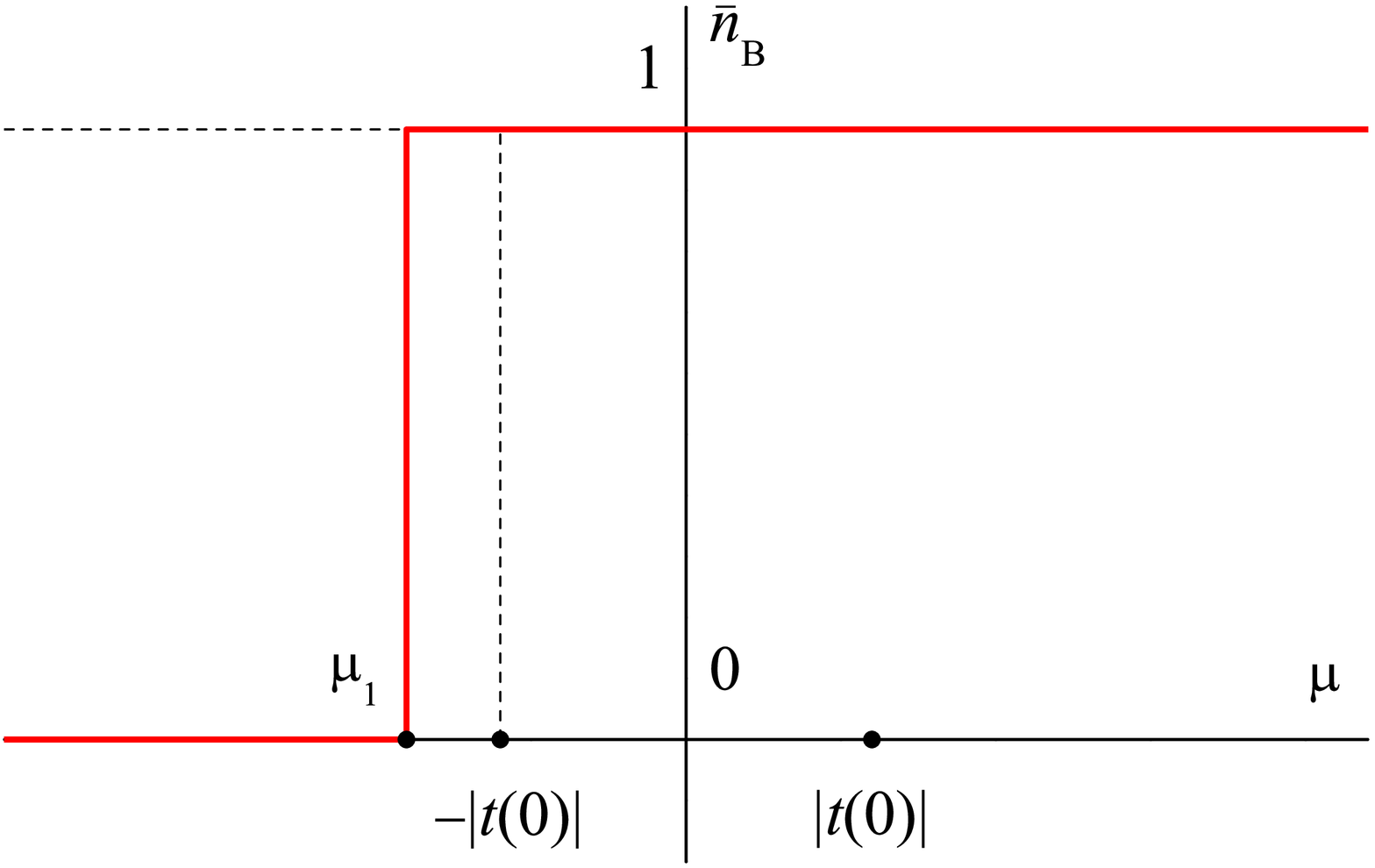}%
\hfill%
\includegraphics[width=0.48\textwidth]{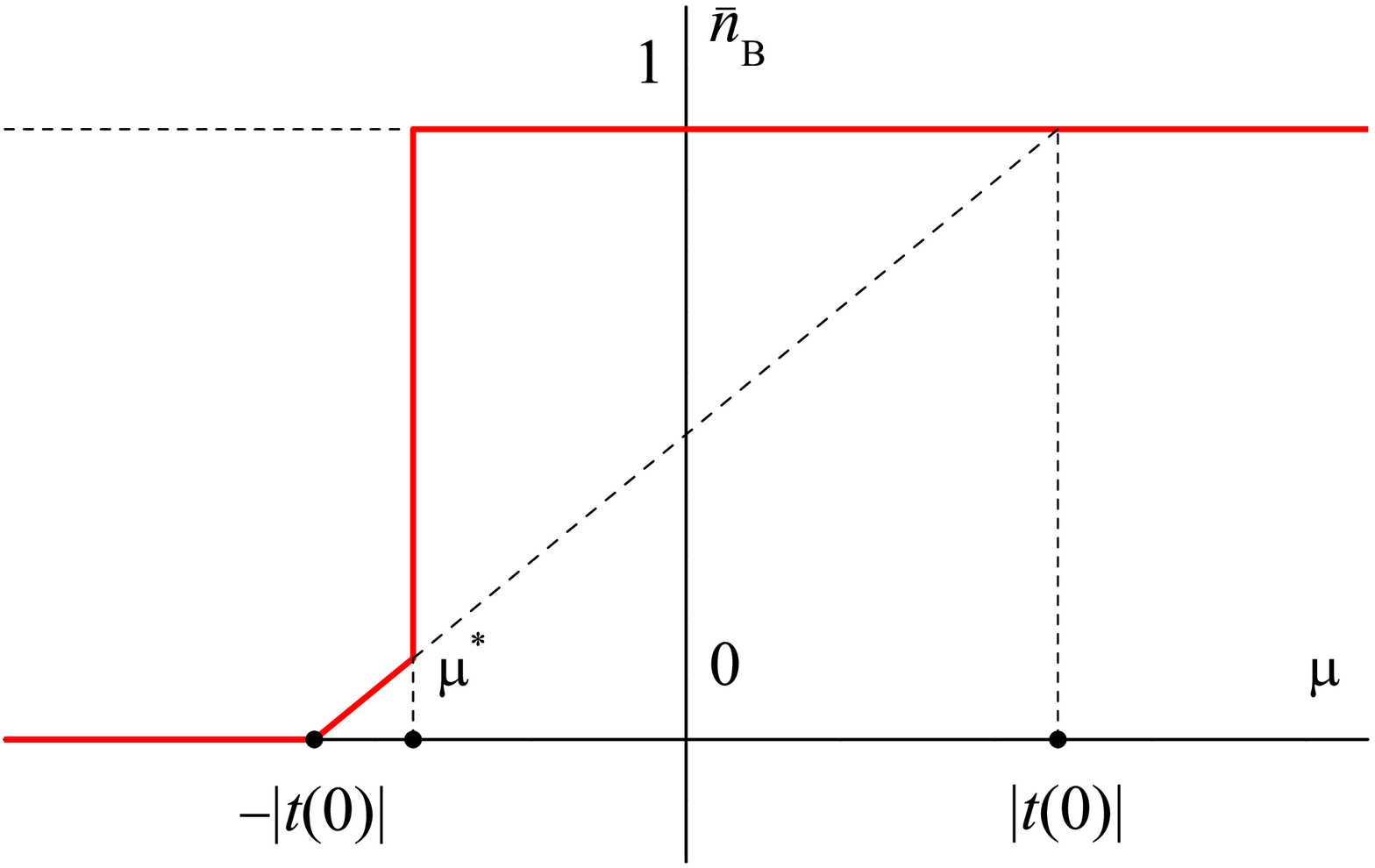}%
\caption{(Colour online) The bosonic occupation number $\bar{n}_{\text{B}}$ as function of the chemical potential of bosons $\mu$ at $T=0$. Left-hand and right-hand graphs correspond to the respective plots in figure~\ref{fig07}.}
\label{fig08}
\end{figure}

The second case [when condition \eqref{istaeq4.1} does not hold] corresponds to large values of bosonic transfer $|t(0)|$. As one can see in figure~\ref{fig07}~(b), minimum values of potential $\Omega$ are reached on three branches in respective intervals: (1) the branch $\Omega_{0}=0$ at $\mu<-|t(0)|$ (the NO phase), (2) the branch $\Omega_{\xi}$ at $-|t(0)|<\mu<\mu^{*}$ (the SF phase with $\xi\neq0$) and (3) the branch $\Omega_{\rho}$ at $\mu>\mu^{*}$ (the phase with $\rho\neq0$). The intersection of branches $\Omega_{\xi}$ and $\Omega_{\rho}$ occurs at the chemical potential $\mu^{*}$ satisfying the following equation
\begin{equation}
	-\frac{1}{4|t(0)|}\left[\mu+|t(0)|\right]^{2}
	=
	-\frac{(W-\delta/2)^{2}}{2W}-\mu
	,
\label{istaeq4.2}
\end{equation}
which gives
\begin{equation}
	\mu^{*}
	=
	-|t(0)|-\sqrt{\frac{2|t(0)|}{W}}\left(W-\delta/2\right)
	.
\label{istaeq4.3}
\end{equation}
In this case, the SF phase exists at the rise of $\mu$ as an intermediate one prior to the phase with $\rho\neq0$. The phase transition to the SF phase at $\mu=-|t(0)|$ is of the second order while the phase transition at $\mu=\mu^{*}$ between the phases with non-zero $\xi$ and $\rho$ is of the first order.
In a similar way, if condition~\eqref{istaeq4.1} is fulfilled and the SF phase is absent, then the transition to the $\rho$ phase is of the first order. 

The behaviour of the boson concentration $\bar{n}_{\text B}$ in the region of these phase transitions is presented in figures~\ref{fig08}~(a) and \ref{fig08}~(b). The phase transition of the first order to the $\rho$ phase is accompanied by a steplike change of $\bar{n}_{\text B}$, while  $\bar{n}_{\text B}(\mu)$ is a linear function in the SF phase. Such a dependence on $\mu$ at $T=0$ is characteristic of the SF phase of the BHM \cite{Sheshadri:93:257,Pai:08:014503}.

\begin{figure}[!t]
\includegraphics[width=0.48\textwidth]{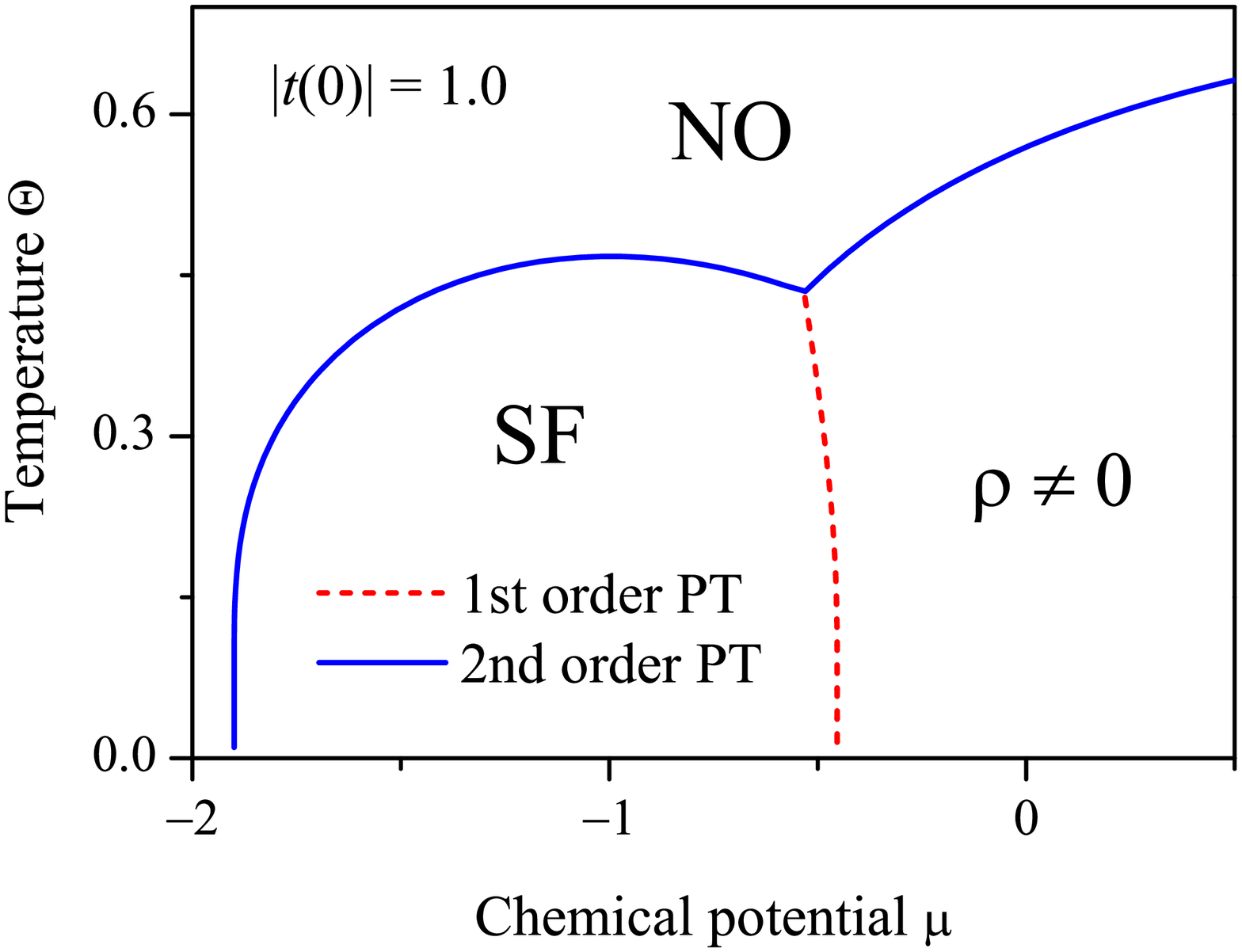}%
\hfill%
\includegraphics[width=0.48\textwidth]{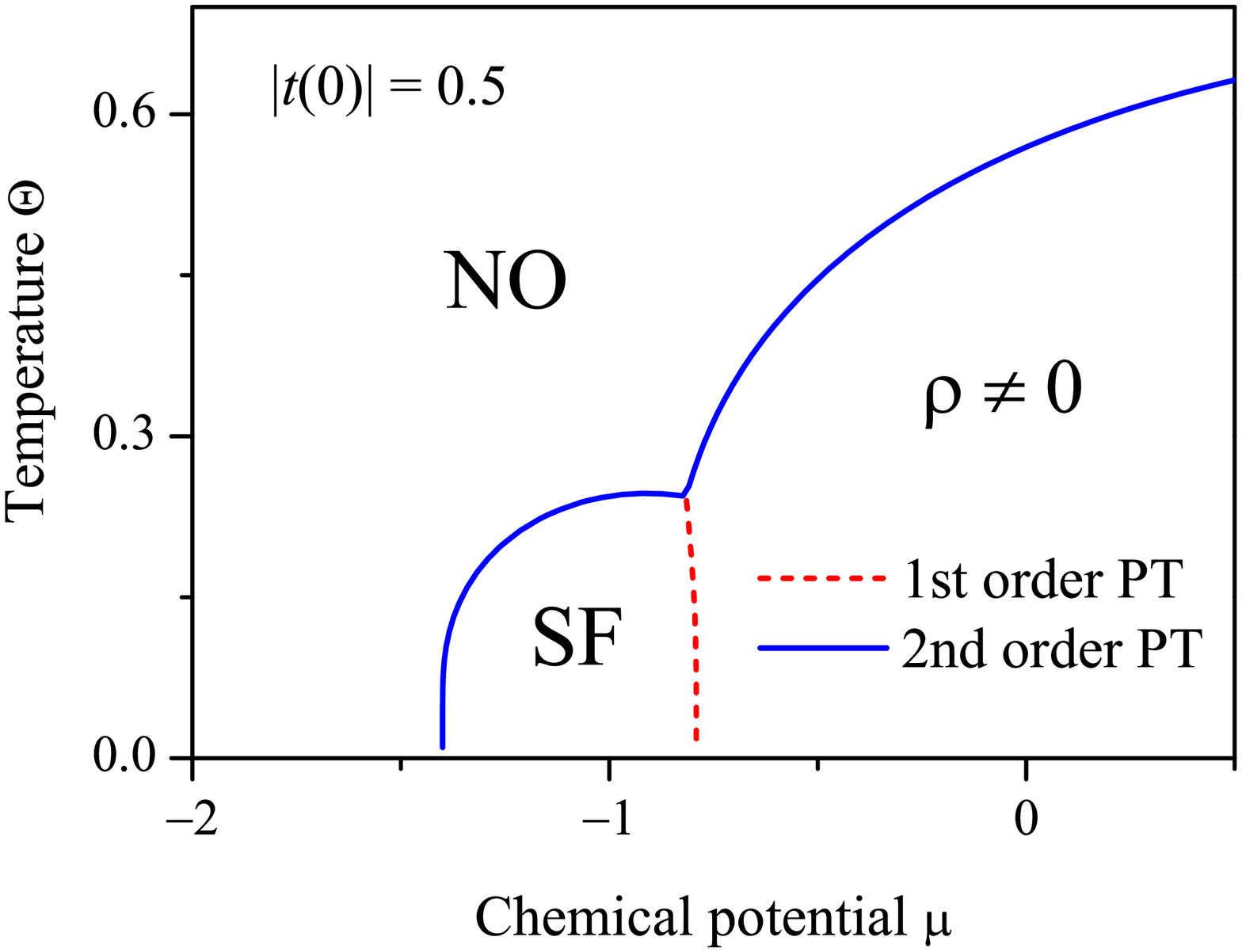}%
\\ [1.4ex]
\includegraphics[width=0.48\textwidth]{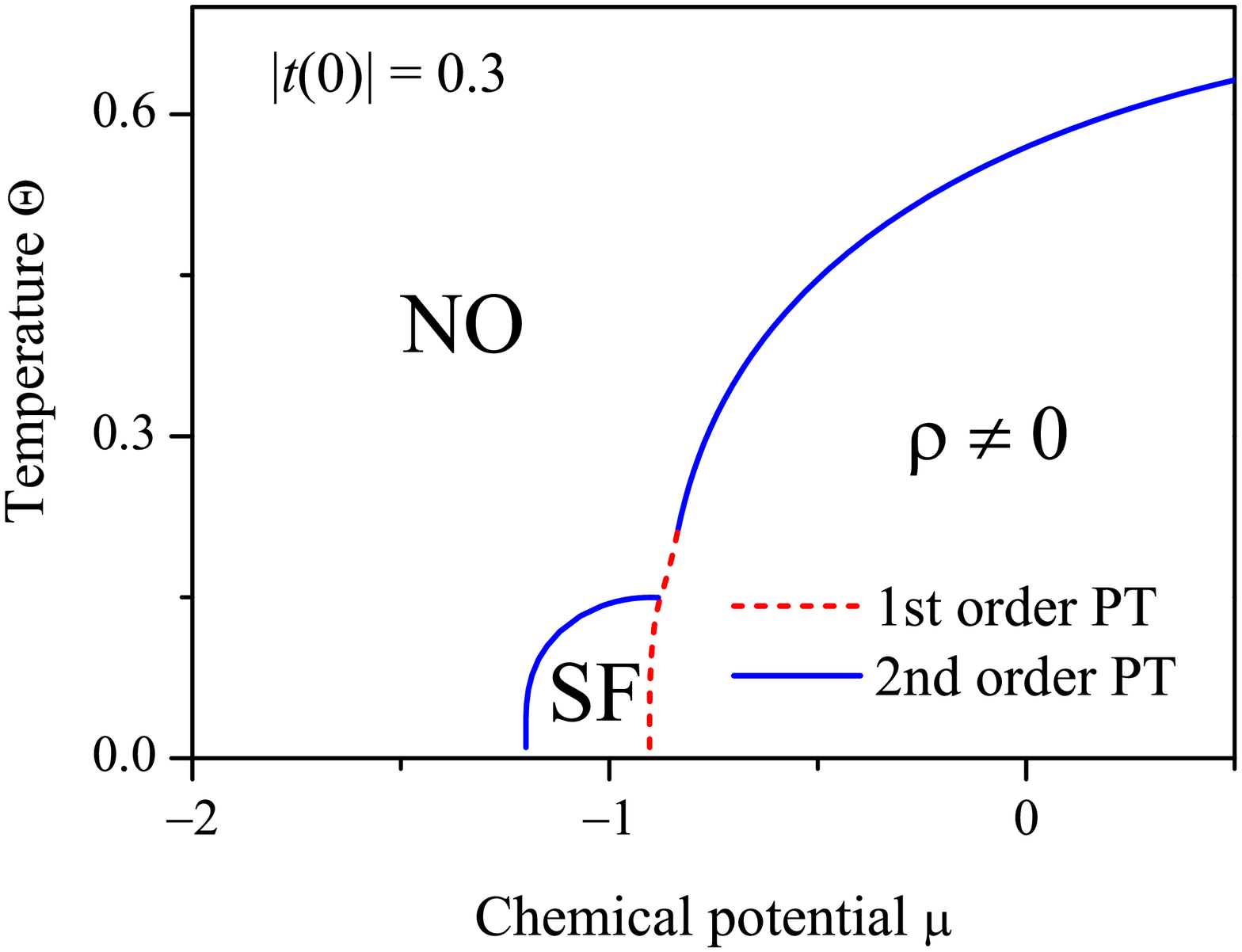}%
\hfill%
\includegraphics[width=0.48\textwidth]{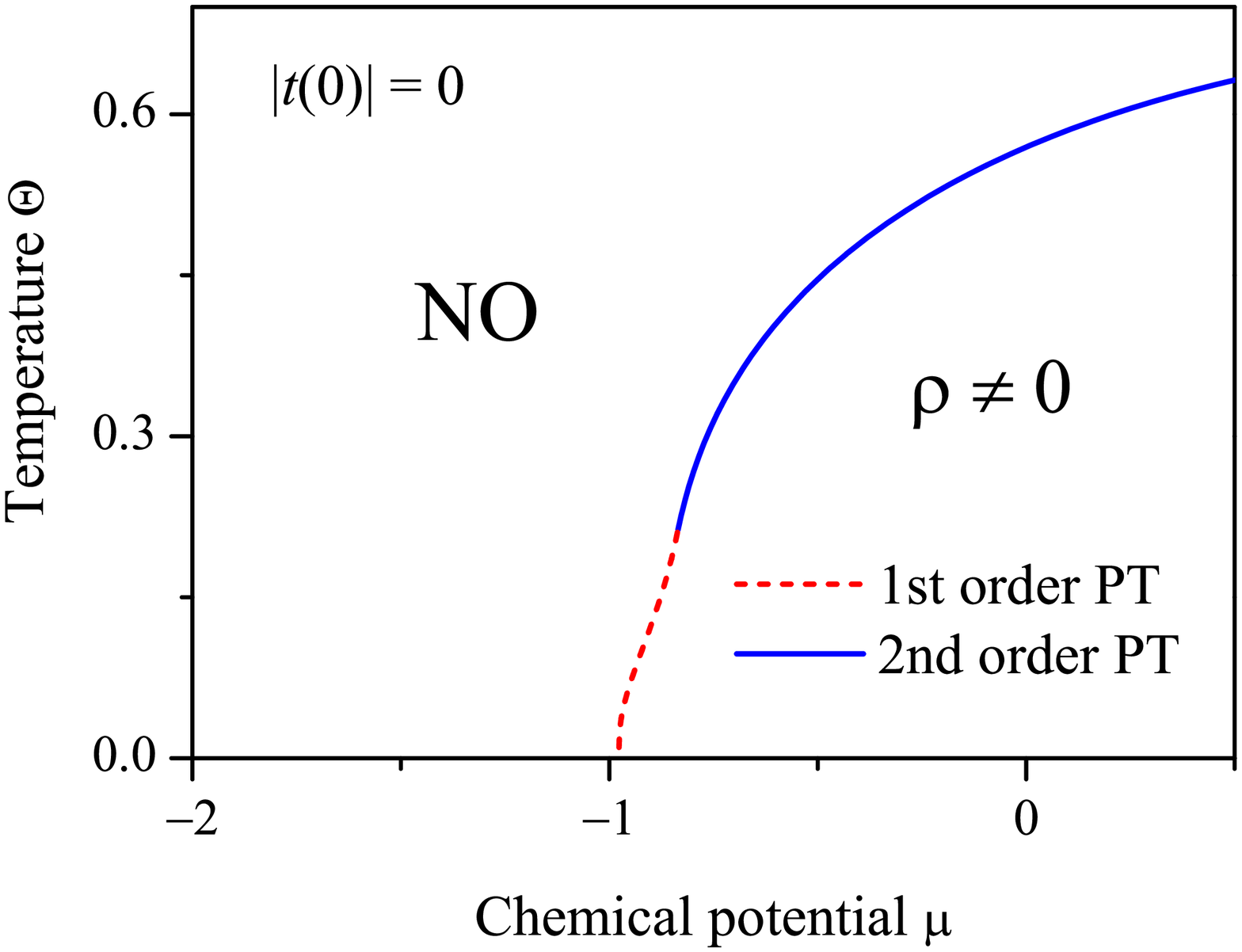}%
\\ [1.4ex]
\includegraphics[width=0.48\textwidth]{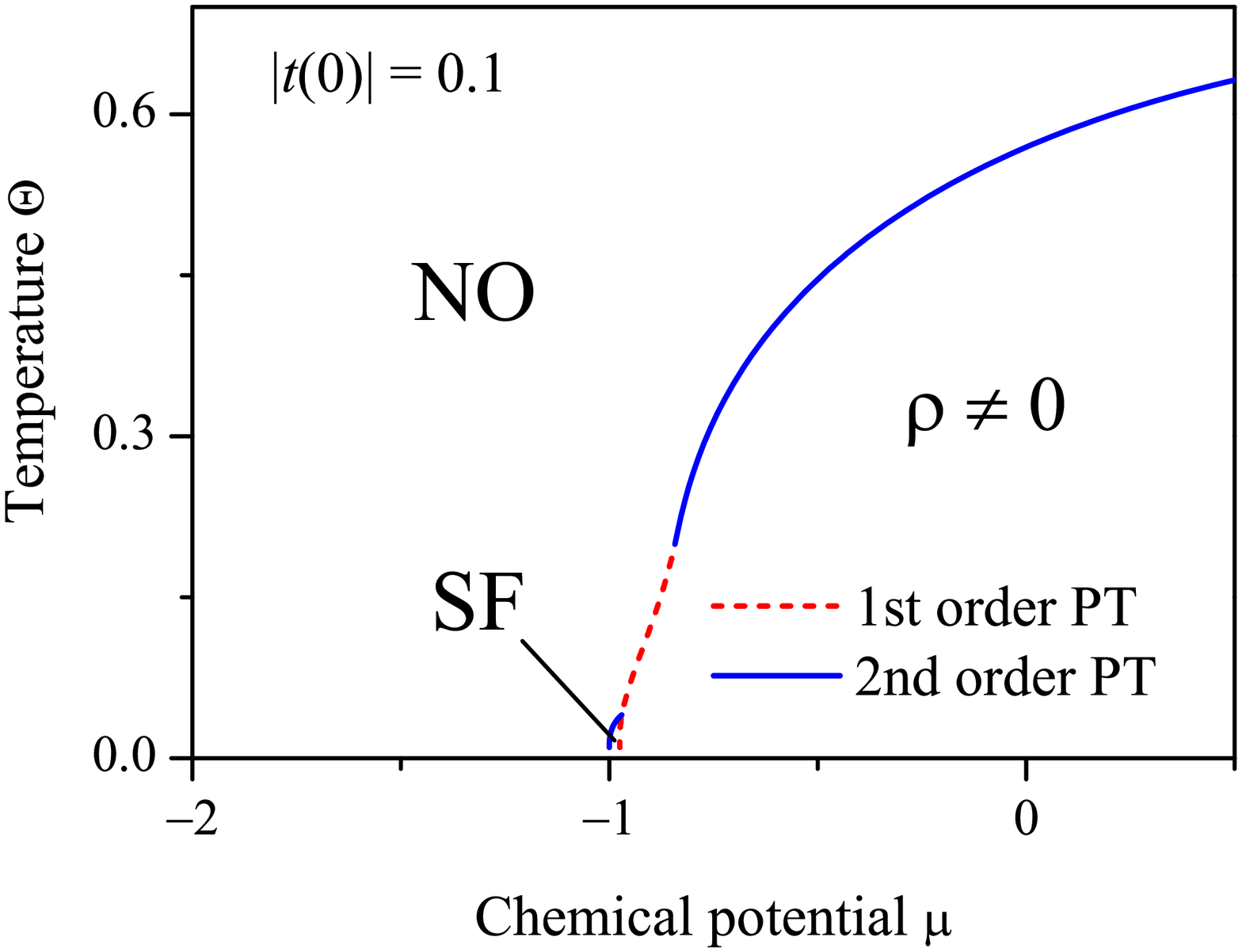}%
\hfill%
\quad\;\includegraphics[width=0.51\textwidth]{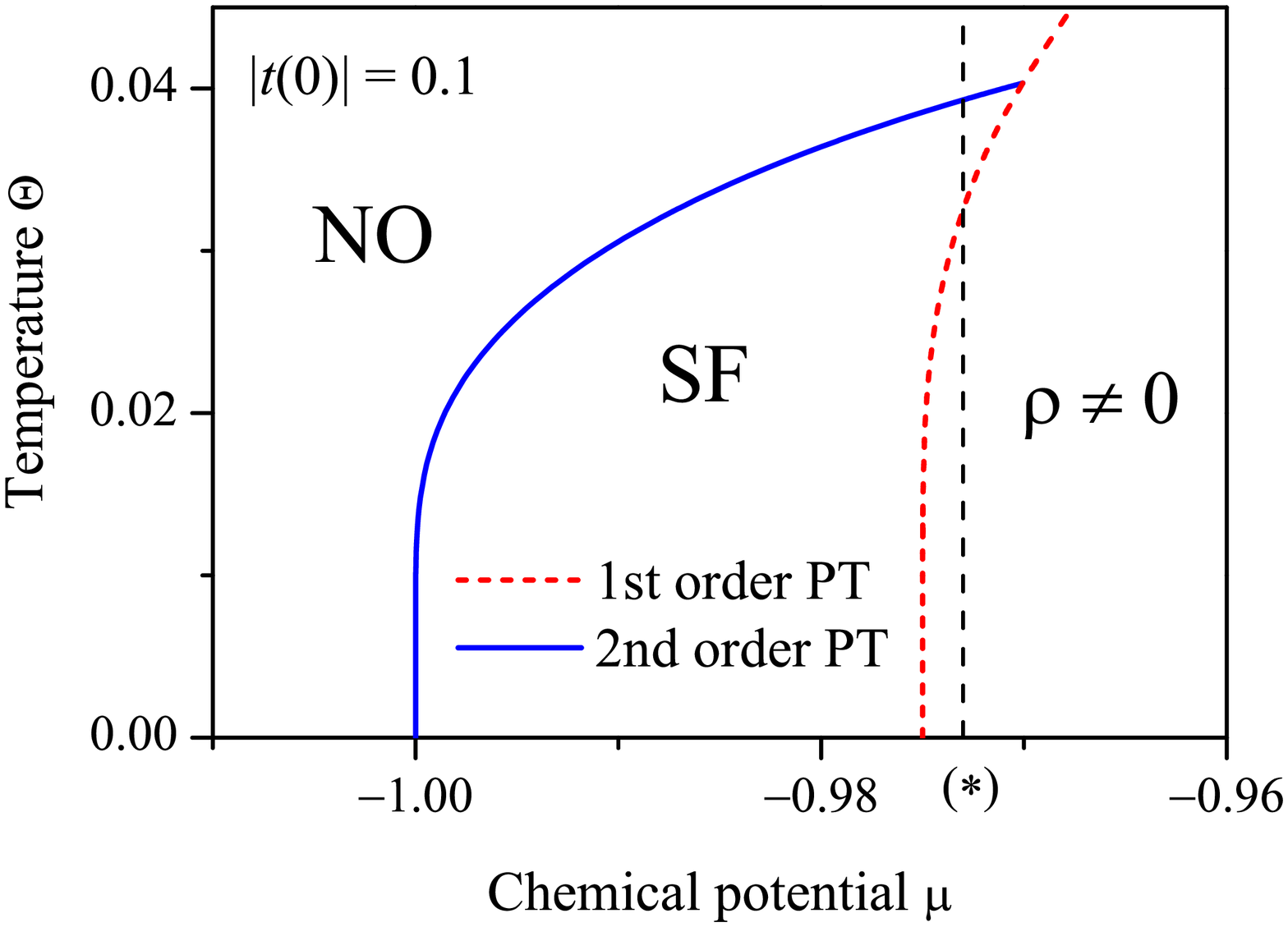}%
\caption{(Colour online) Phase diagrams $(T,\mu)$: regions of the NO phase and phases with $\xi\neq0$ and $\rho\neq0$ at various values of $|t(0)|$ ($\delta=0.9$, $W=0.8$). Designations of phase transition lines are identical to the ones in figure~\ref{fig04}. The possibility of the SF phase existence as an intermediate one at the rise of temperature [the vertical line marked (*)] is shown on the last graph.}
\label{fig09}
\end{figure}

\begin{figure}[!t]
\centerline{\includegraphics[width=0.6\textwidth]{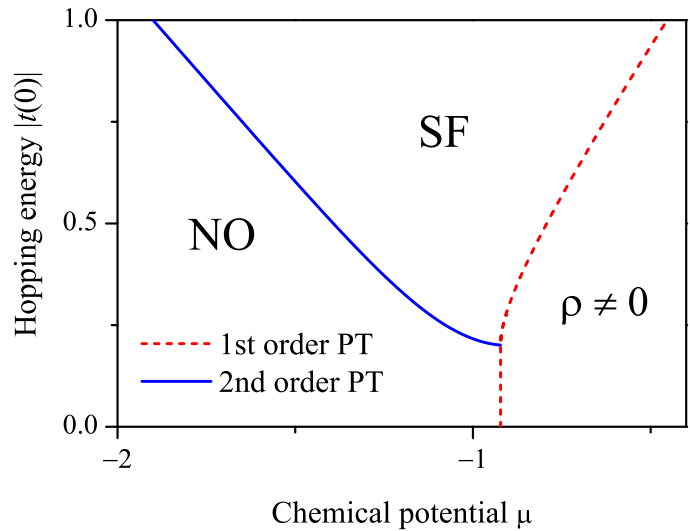}}
\caption{(Colour online) Phase diagram $(|t(0)|,\mu)$ at $T=0.1$, $\delta=0.9$ and $W=0.8$. Designations of phase transition lines are identical to the ones in figure~\ref{fig04}.}
\label{fig10}
\end{figure}

The above considerations are valid at the absolute zero temperature. Numerical calculations prove that at $T\neq0$, the SF phase remains intermediate as in the above discussed case. Phase diagrams $(T,\mu)$ at various values of parameter $|t(0)|$ (presented in figure~\ref{fig09}) demonstrate a narrowing of the SF region at a decrease of $|t(0)|$. The achievable maximum temperature of the SF phase also decreases. At $|t_{0}|<|t_{0}|_{\text{cr}}$, the SF phase is absent ($|t_{0}|_{\text{cr}} \approx 0.2$ at $T=0.1$, $\delta=-0.9$, $W=0.8$; see the phase diagram ($|t_{0}|,\mu$) in figure~\ref{fig10}).

At $T\neq0$, the phase transition between the $\xi$ and $\rho$ phases is also of the first order. At small values of $|t(0)|$, the phase transition between NO and $\rho$ phases changes its order from the first to the second one at the rise of temperature (figure~\ref{fig09}). At large enough values of $|t(0)|$, this transition is of the second order starting from the triple point where all three phases are in equilibrium.

\section{Conclusions}

Thermodynamics of the lattice bosonic system, studied within the framework of the two-state HCB model, is affected by an intersite interaction, caused by displacements of particles from their equilibrium positions on lattice sites, which can lead to spontaneous symmetry breaking and to the appearance of a phase with uniform or space-modulated displacements. Matrix elements of displacements are related to bosonic transitions between the ground and excited states. Thus, the mentioned orderings of displacements are analogous to dipole orderings in ferroelectrics. An intersite particle transfer leads to the instability with respect to the BE condensation that exists in such intervals of the chemical potential of bosons where their concentration is non-integer and varies. Competition between this effect and the tendency to freezing of the displacements affects the shape of phase diagrams defining the regions of the NO phase, the SF phase ($\xi\neq0$) and the phase with spontaneous displacements ($\rho\neq0$). The phase with non-zero values of both order parameters $\rho$ and $\xi$ is unstable, as proved by a numerical analysis of the $\Omega(\mu)$ dependence. Such a state could be a supersolid phase which is well known from the theory of phase transitions in optical lattices derived within the framework of an extended Bose-Hubbard model with frustration (see, for example, \cite{Zhang:13:174515}).

In particular, our model approach is aimed at optical lattices formed by the system of local double wells. In this case, a decisive contribution is made by the intersite hopping of particles residing in the ground vibrational state. Hence, a ``frozen'' displacement (when $\rho\neq0$) corresponds to localization in  certain minima of the double wells.  Such a phenomenon is similar to the ordering of protons in hydrogen bonds in hydrogen-bonded ferroelectric crystals. However, there is a difference: the number of protons in these crystals is constant (one proton per bond). The exception is a crystal with a superprotonic (superionic) phase where the number of virtual H-bonds exceeds the number of protons, so the average bond occupation is fractional ($\bar{n}_{i}<1$).
The crystals of the M$_3$H(XO$_4$)$_2$ family (where M${}={}$Rb, Cs, NH$_4$; X${}={}$S, Se), in which in the superprotonic phase $\bar{n}_i=1/3$, are an example, see~\cite{Stasyuk:97:135}.
Thus, our model can also be applied to a description of phases with a superprotonic conductivity existing at temperatures higher than the ones characteristic of the ordered phases with an integer occupation per hydrogen bond. A possibility of the above-mentioned scenario is illustrated in figure~\ref{fig09}, 
where at some relations between the model parameters and certain values of the boson chemical potential, the SF phase can appear at intermediate temperatures.

It should be mentioned that the model of the \eqref{istaeq1.1} type, describing the proton ordering as well as superionic transition as a disordering process, was proposed for the mentioned class of crystals and was used for the calculation of proton conductivity \cite{Stasyuk:97:135,Pavlenko:01:4607}. 
Anyway, this issue calls for a separate and more detailed study.

\appendix

\section{Optical lattice with double-well potentials}%
\label{app-1}

\begin{figure}[!t]
\centerline{\includegraphics[width=0.6\textwidth]{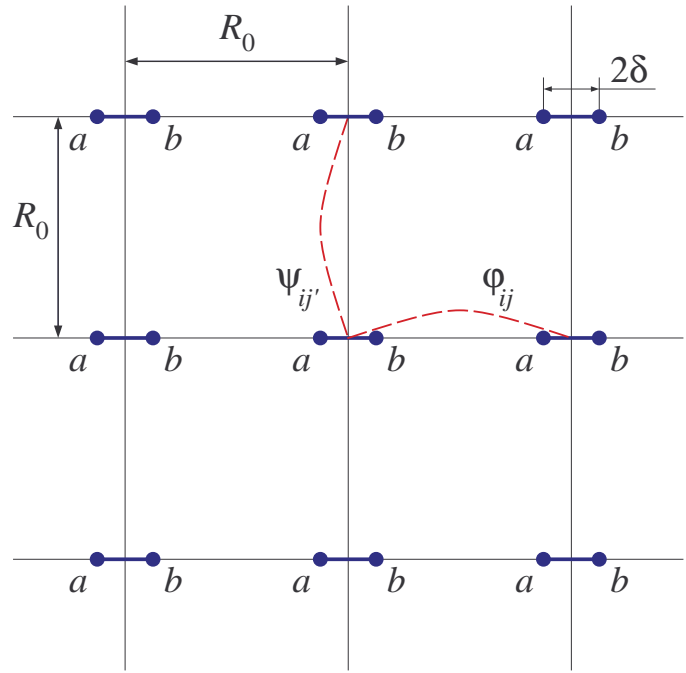}}
\caption{(Colour online) Interaction of bosons in a lattice with  local double-well potentials.}
\label{fig11}
\end{figure}

As an example, let us consider an optical lattice with local double-well potentials. For simplicity, the further consideration will be limited to the case of a simple cubic lattice with the lattice constant~$R_{0}$ where double wells are oriented along a certain axis (figure~\ref{fig11}). As usual, we assume that bosons residing in the lattice potential wells are localized in vibrational states with the lowest energy. For the lattice site~$i$, these states $|i,a\rangle$ and $|i,b\rangle$ are localized in minima $a$ and $b$, respectively.

We take into account the hopping of bosons between the nearest positions in the lattice: (1) hopping between the minima $a$ and $b$ in a single double well and (2) hopping between neighbour double wells \cite{Trotzky:08:295}. In the first case, the Hamiltonian for the well on the lattice site $i$ looks as follows:  
\begin{equation}
	\hat{H}_{0}
	=
	\Omega(c_{ia}^{+} c_{ib}+c_{ib}^{+}c_{ia})
	-
	\mu(c_{ia}^{+}c_{ia}+c_{ib}^{+}c_{ib})
	,
\label{istaeqD1}
\end{equation}
where $c_{ia}(c_{ia}^{+})$ and $c_{ib}(c_{ib}^{+})$ are operators of destruction (creation) of bosons in potential minima $|i,a\rangle$ and $|i,b\rangle$, $\Omega$ is the parameter describing the hopping (tunnelling) between them, $\mu$ is the chemical potential of bosons. By means of transformation
\begin{align}
	c_{ia}
	&=
	\frac{1}{\sqrt{2}}(d_{i1}+d_{i2}),
	\notag\\
	c_{ib}
	&=
	\frac{1}{\sqrt{2}}(d_{i1}-d_{i2}),
\label{istaeqD2}
\end{align}
which is equivalent to the transition from localized states
$|i,a\rangle$ and $|i,b\rangle$ to their symmetric
\[
	\frac{1}{\sqrt{2}}\left(|i,a\rangle+|i,b\rangle\right)\equiv|i,1\rangle
\]
and asymmetric
\[
	\frac{1}{\sqrt{2}}\left(|i,a\rangle-|i,b\rangle\right)\equiv|i,2\rangle
\]
combinations, Hamiltonian \eqref{istaeqD1} becomes diagonal:
\begin{equation}
	\hat{H}_{i}=E_{1}d_{i1}^{+}d_{i1}+E_{2}d_{i2}^{+}d_{i2}\,,
\label{istaeqD3}
\end{equation}
where $E_{1,2}=\pm\Omega-\mu$. At $\Omega<0$, the ground state is a symmetric one $|i,1\rangle$. Due to tunnelling, the excited state is separated from the ground one by a gap of the width $2|\Omega|$.

Hopping between adjacent double wells (the second case) is described by the operator
\begin{equation}
	\hat{H}
	=
	\sum_{ij}\nolimits^{(\|)}\sum_{\alpha\beta}
		\varphi_{ij}^{\alpha\beta}c_{i\alpha}^{+}c_{j\beta}
	+
	\sum_{ij}\nolimits^{(\perp)}\sum_{\alpha\beta}
		\Psi_{ij}^{\alpha\beta}c_{i\alpha}^{+}c_{j\beta}
	\,,
\label{istaeqD4}
\end{equation}
where the first and the second terms correspond to a bosonic transfer in the longitudinal (parallel to the orientation of double wells) and transverse directions, respectively ($\alpha,\beta=a,b$).

When the distance $r_{ab}$ between positions $(a)$ and $(b)$ in a potential well is much smaller compared to the distance $R_{0}$ between the neighbour wells \cite{Trotzky:08:295}, one can perform expansions in terms of the ratio $r_{ab}/R_{0}$. In this way, the hopping parameters are expressed as follows:
\begin{align}
	&
	\varphi_{ij}^{aa}
	=
	\varphi_{ij}^{bb}=\varphi(R_{0}), 
	&
	&
	\varphi_{ij}^{ab}
	=
	\varphi_{ij}^{ba}=\varphi(R_{0})\pm\varphi'(R_{0})\cdot 2r_{ab}
	;
	\notag\\
	&
	\psi_{ij}^{aa}
	=
	\psi_{ij}^{bb}=\psi(R_{0}), 
	&
	&
	\psi_{ij}^{ab}
	=
	\psi_{ij}^{ba}=\psi(R_{0})+\psi'(R_{0})\cdot \frac{2r_{ab}^{2}}{R_{0}}
	.
\label{istaeqD5}
\end{align}

In terms of operators $d_{i1}$ and $d_{i2}$, the Hamiltonian \eqref{istaeqD4} written down as $\hat{H}=\hat{H}_{\|}+\hat{H}_{\perp}$, looks as follows:
\begin{align}
	\hat{H}_{\|}
	&=
	\sum_{i<j}\nolimits^{(\|)}
	\left[
		2\varphi(R_{0})d_{i1}^{+}d_{j1}-\varphi'(R_{0})2r_{ab}\cdot d_{i1}^{+}d_{j2}
	+
		\varphi'(R_{0})2r_{ab}\cdot d_{i2}^{+}d_{j1}
	\right]
	+
	{\text{h.c.}}\,,
\label{istaeqD6}
	\\
	\hat{H}_{\perp}
	&=
	\sum_{ij}\nolimits^{(\perp)} 
	\left\{
		\left[2\psi(R_{0})+\psi'(R_{0})\cdot \frac{2r_{ab}^{2}}{R_{0}}\right]
		d_{i1}^{+}d_{j1}
	-
		\psi'(R_{0})\cdot \frac{2r_{ab}^{2}}{R_{0}}d_{i2}^{+}d_{j2}
	\right\}
	+
	{\text{h.c.}}
	\,.
\label{istaeqD7}
\end{align}

At a small ratio $r_{ab}/R_{0}$, the major contribution is made by hopping between symmetric states. The corrections start in the MFA from the second order (linear corrections mutually compensate). Neglecting the terms of the second (and higher) order, the bosonic transfer is described by the Hamiltonian
\begin{equation}
	\hat{H}_{\|}+\hat{H}_{\perp}
	=
	\sum_{ij}\nolimits^{(\|)}t_{ij}^{\|}d_{i1}^{+}d_{j1}
	+
	\sum_{ij}\nolimits^{(\perp)}t_{ij}^{+}d_{i1}^{+}d_{j1}
	.
\label{istaeqD8}
\end{equation}

Let us take $t_{ij}^{\|}=t_{ij}^{\perp}\equiv t_{ij}$, having in mind that the effects of local anisotropy are negligibly small $[\varphi(R_{0})\approx\psi(R_{0})]$. As a result, we obtain the Hamiltonian of a two-state model
\begin{align}
	\hat{H}
	&=
	\Omega\sum_{i}\left(d_{i1}^{+}d_{i1}-d_{i2}^{+}d_{i2}\right)
	-
	\mu\sum_{i}\left(d_{i1}^{+}d_{i1}+d_{i2}^{+}d_{i2}\right)
	+
	\sum_{ij}t_{ij}d_{i1}^{+}d_{j1}
	\,,
\label{istaeqD9}
\end{align}
where the transfer occurs over the ground states, and $a-2|\Omega|=\delta$ corresponds to the excitation energy (the energy difference between the excited and ground states of a site). After the substitution $\mu\rightarrow\mu-\delta/2$, operator \eqref{istaeqD8} transforms to Hamiltonian \eqref{istaeq1.1}.

\section{State with modulation of displacements}%
\label{app-2}

To describe the modulation with a doubled lattice constant, it is convenient to divide the initial lattice into two sublattices. Let us put
\begin{equation}
	\langle\hat{x}_{i1}\rangle
	=
	\rho_{1}\,,
	\qquad 
	\langle\hat{x}_{i2}\rangle
	=
	\rho_{2}
\label{istaeqD10}
\end{equation}
for the first and second of them, respectively. If the intersite interaction between displacements $\Phi_{ij}$ is considered in the MFA, an external field is absent and $t_{ij}\rightarrow0$, then the Hamiltonian can be written down as follows:
\begin{equation}
	\hat{H}
	=
	\sum_{i_{1}}\hat{H}_{i_1}
	+
	\sum_{i_2}\hat{H}_{i_2}
	-
	\frac{N}{2}\phi(0)\rho_{1}\rho_{2}\,,
\label{istaeqD11}
\end{equation}
where
\begin{align}
	\hat{H}_{i_{1}}
	&=
	-
	\mu X_{i_{1}}^{11}
	+
	(\delta-\mu)X_{i_{1}}^{22}
	+
	\rho_{2}\Phi(0)d\left(X_{i_{1}}^{12}
	+
	X_{i_{1}}^{21}\right),
	\notag\\
	\hat{H}_{i_{2}}
	&=
	-
	\mu X_{i_{2}}^{11}
	+
	(\delta-\mu)X_{i_{2}}^{22}
	+
	\rho_{1}\Phi(0)d\left(X_{i_{2}}^{12}
	+
	X_{i_{2}}^{21}\right)
	.
\label{istaeqD12}
\end{align}
Eigenvalues of operators \eqref{istaeqD12} acting on the basis of states $|0\rangle$, $|1\rangle$ and $|2\rangle$ look as follows:
\begin{align}
	&
	\lambda_{1}^{(1,2)}
	=
	0, 
	\quad  \lambda_{2}^{(1,2)}
	=
	\frac{\delta}{2}-\mu+\sqrt{\delta^{2}/4+B_{(2,1)}^{2}}\,,
	\notag\\
	&
	\lambda_{3}^{(1,2)}
	=
	\frac{\delta}{2}-\mu-\sqrt{\delta^{2}/4+B_{(2,1)}^{2}}\,,
\label{istaeqD13}
\end{align}
with
\begin{equation}
	B_{1}=\rho_{1}\Phi(0)d,
	\qquad 
	B_{2}=\rho_{2}\Phi(0)d.
\label{istaeqD14}
\end{equation}
In this case, the grand canonical potential is given by the expression
\begin{align}
	\Omega/N
	&=
	-
	\frac12\Phi(0)\rho_{1}\rho_{2}
	-
	\frac{\Theta}{2}
	\left[
		1
		+
		2\cosh\beta\sqrt{\delta^{2}/4+B_{1}^{2}}
			\cdot
			{\rm e}^{-\beta(\frac{\delta}{2}-\mu)}
	\right]
	\notag\\
	&\quad
	-
	\frac{\Theta}{2}
	\ln
	\left[
		1
		+
		2\cosh\beta\sqrt{\delta^{2}/4+B_{2}^{2}}
			\cdot
			{\rm e}^{-\beta(\frac{\delta}{2}-\mu)}
	\right]
\label{istaeqD15}
\end{align}
(it is taken into account that the number of sites in the sublattice is equal to $N/2$).

Conditions of equilibrium 
\begin{equation}
\frac{\partial\Omega/N}{\partial\rho_{1}}=0, \qquad \frac{\partial\Omega/N}{\partial\rho_{2}}=0
\label{istaeqD16}
\end{equation}
define the equations for the order parameters $\rho_{1}$ and $\rho_{2}$
\begin{equation}
	\rho_{1(2)}
	=
	-
	\frac{2\sinh\beta\sqrt{\delta^{2}/4+B_{2(1)}^{2}}}
		{{\rm e}^{\beta(\frac{\delta}{2}-\mu)}
	+
	2\cosh\beta\sqrt{\delta^{2}/4+B_{2(1)}^{2}}}
		\cdot
		\frac{d^{2}\phi(0)}{\sqrt{\delta^{2}/4+B_{2(1)}}}
		\cdot
		\rho_{2(1)}
	.
\label{istaeqD17}
\end{equation}
When $\Phi_{ij}$ is different from zero only for nearest neighbours, its Fourier transform $\Phi\,(\,\vec{q}\,)$ has extrema only in the centre and on the boundary of the Brillouin zone. Hence, there are two possible cases.
\begin{enumerate}
\item 
$\rho_{1}=\rho_{2}=\rho$ (a uniform ordering of displacements). Equation \eqref{istaeqD17} reduces to the above-written one \eqref{istaeq3.3} with non-zero solutions at $\Phi(0)<0$.
\item 
$\rho_{1}=-\rho_{2}=\rho$ (an alternating modulation with a doubled lattice constant). Here, equation \eqref{istaeqD17} should be rewritten for $\rho$ with a plus sign of the right-hand side. Non-zero solutions exist at $\Phi(0)>0$.
\end{enumerate}

\ukrainianpart

\title{Бозе-ейнштейнівська конденсація та/або модуляція ``зміщень'' у двостановій моделі Бозе-Хаббарда}
\author{І.В. Стасюк, О.В. Величко}
\address{Інститут фізики конденсованих систем НАН України, вул. Свєнціцького, 1, 79011 Львів, Україна}

\makeukrtitle

\begin{abstract}
\tolerance=3000%
Досліджено нестійкості, результатом яких є бозе-конденсація та/або модуляція ``зміщень'' у системі квантових частинок, які описуються двостановою моделлю Бозе-Хаббарда з врахуванням взаємодії між зміщеннями частинок у різних вузлах ґратки. Вивчено можливість модуляції з подвоєнням періоду ґратки, а також однорідного зміщення частинок з рівноважних позицій. Проаналізовано умови реалізації згаданих нестійкостей та фазових переходів до SF фази та ``впорядкованої'' фази (зі замороженими зміщеннями), вивчено поведінку параметрів порядку та побудовано фазові діаграми --- як аналітично (для основного стану), так і числовим способом (при ненульовій температурі). Показано, що SF фаза може з'явитися як проміжна між нормальною та ``впорядкованою'' фазами, в той час як фаза ``суперсолід'' є термодинамічно нестійкою і не реалізується. Обговорено застосовність отриманих результатів до ґраток з локальними потенціалами типу подвійних ям.
\keywords модель Бозе-Хаббарда, жорсткі бозони, бозе-конденсат, фазовий перехід, зміщення частинок, однорідне та модульоване впорядкування

\end{abstract}

\end{document}